\documentclass[aps,  prb,twocolumn,
superscriptaddress,  floatfix,  10pt   
]{revtex4-2}
\usepackage{float}
\usepackage{graphicx}
\usepackage{amsmath,amssymb}
\usepackage{tikz}
\usepackage{braket}
\usepackage{pgfplots}
\usepackage{bm}\pgfplotsset{compat=1.17}
\usepackage[version=4]{mhchem}
\usepackage{physics}

\usepackage{enumitem}
\usepackage{mathrsfs}
\usepackage{subcaption}

\usetikzlibrary{
	shapes.geometric,
	shadows,
	patterns,
	perspective,
	decorations,
	decorations.text
}

\usetikzlibrary{decorations.markings,calc,decorations.pathreplacing}

\usepackage{xcolor}

\definecolor{linkcolor}{RGB}{0,0,255} 
\definecolor{citecolor}{RGB}{10,155,55}    
\definecolor{urlcolor}{RGB}{0,0,255}

\definecolor{dy}{rgb}{0.9,0.9,0.4}
\definecolor{dr}{rgb}{0.95,0.65,0.55}
\definecolor{db}{rgb}{0.5,0.8,0.9}
\definecolor{dg}{rgb}{0.2,0.9,0.6}
\definecolor{DarkGreen}{rgb}{0.0,0.5,0.0}
\definecolor{BrickRed}{rgb}{0.2,0.2,0.8}
\definecolor{Navy}{rgb}{0.8,0.2,0.1}
\definecolor{phaseK}{HTML}{CC79A7}
\definecolor{phaseY}{HTML}{56B4E9}
\definecolor{phaseU}{HTML}{E69F00}
\usepackage{hyperref}
\hypersetup{
	colorlinks   = true,
	linkcolor    = linkcolor,
	citecolor    = citecolor,
	urlcolor     = urlcolor,
	linktoc      = all,
	pdfborder    = {0 0 0},
}

\begin{document}
	
	\title{Thermodynamics in a split Hilbert space:  Quantum impurity at the edge of a one-dimensional superconductor}
	\author{Pradip Kattel}
	\email{pradip.kattel@rutgers.edu}
	
	\author{Abay Zhakenov}
	\author{Natan Andrei}
	\affiliation{Department of Physics and Astronomy, Center for Materials Theory, Rutgers University, Piscataway, New Jersey 08854, USA
	}
	
	\begin{abstract}
We present a thermodynamic description of a single magnetic impurity at the edge of a superconducting wire. 
The impurity exhibits four phases—Kondo, Yu--Shiba--Rusinov (YSR) I and II, and local moment— a phase diagram richer than in the gapless case, contrary to the expectation that the effects of impurities in gapped hosts are less consequential. We derive the impurity contribution to free energy $F_{\rm imp}(T)$ and entropy in each phase: in Kondo phase, the entropy flows monotonically from $\ln 2$ (UV) to 0 (IR) with critical exponents same as that of the conventional Kondo model; in YSR phases, thermal activation of a midgap bound state produces entropy overshoots above $\ln 2$, saturating to $\ln 2$ at high $T$ and approaching either 0 or $\ln 2$ at low $T$ depending on whether impurity is screened or not; in the local-moment phase the impurity remains effectively decoupled, with entropy near $\ln 2$, with some intermediate-temperature features that progressively fade as $\delta \to 0$. These behaviors, including the entropy overshoots in the YSR and local-moment phases, stem from a splitting of the Hilbert space into distinct excitation towers: one in the Kondo phase, two in YSR~I, and three in YSR~II and the local-moment phase. Resolving these tower structures and thereby going beyond conventional TBA yields closed-form analytic expressions for the impurity contribution to the free energy and entropy across the entire phase diagram.
    \end{abstract}
	
	\maketitle

The interplay between magnetic impurities and superconducting hosts has been of enduring interest since the seminal works of Yu~\cite{Yu}, Shiba~\cite{Shiba}, and Rusinov~\cite{Rusinov}, which, treating the impurity as a static classical spin, uncovered in-gap Yu–Shiba–Rusinov (YSR) bound states. Fully quantum treatments have since revealed a richer phase structure, including YSR, Kondo-screened, and local-moment (unscreened) phases~\cite{pasnooriKondoSup,pasnoori2022rise,moca2024spectral,moca2021kondo}. Many ground-state properties~\cite{pasnoori2022rise,moca2024spectral}, correlation functions~\cite{moca2021kondo}, and some dynamical features~\cite{wei2025kondo} have been explored. The setup has been extended to higher-spin superconductors, to overscreened Kondo impurities~\cite{kattel2024overscreened} and to SU(3) superconductors~\cite{kattel2025competing}. However, the impurity thermodynamics remained largely unexplored.

\begin{figure}[H]
  \includegraphics[width=\linewidth]{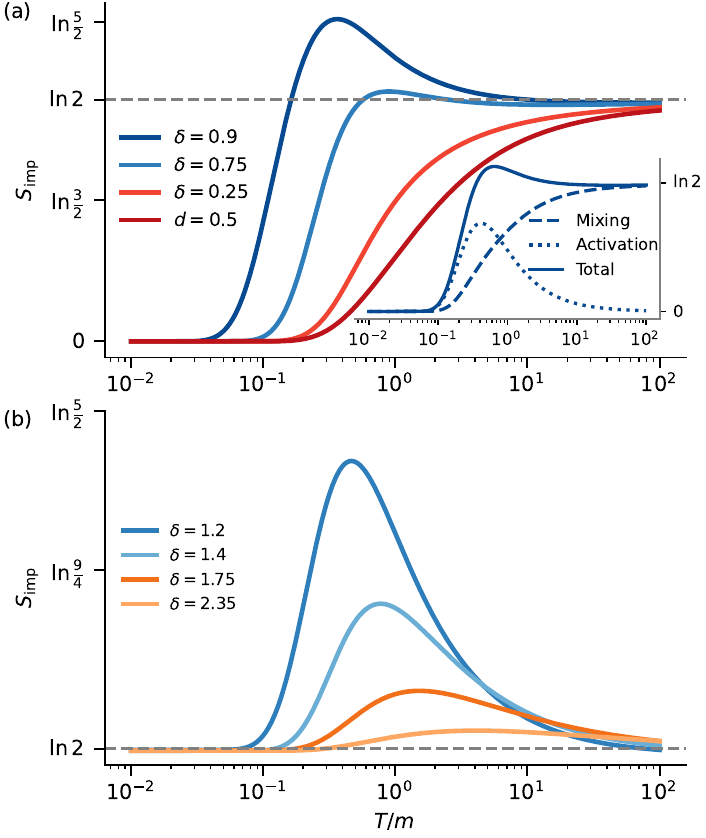}
  \caption{\textbf{Impurity entropy $S_{\mathrm{imp}}(T)$ across phases}. 
    (a) Kondo (two {red shades}, $d=0.5$, $\delta=0.45$) and YSR~I 
    (two {darker blue shades}, $\delta=0.75,0.9$). 
    In the Kondo phase, $S_{\mathrm{imp}}$ increases monotonically from $0$ to $\ln 2$; 
    in YSR~I it is non-monotonic with a bump at $T \sim  m$. 
    Inset: mixing vs.\ activation contributions [Eq.~(11)] for $\delta=0.9$. 
    (b) YSR~II (two lighter {blue shades}, $\delta=1.2,1.4$) and local-moment 
    (two {orange shades}, $\delta=1.75,2.35$). 
    In YSR~II, $S_{\mathrm{imp}}=\ln 2$ at both $T \to 0$ and $T \to \infty$, 
    with a hump at $T \sim  E_\delta$. 
    In the local-moment phase, $S_{\mathrm{imp}}=\ln 2$ in both limits, and bumps 
    fade deeper in the phase, leaving $S_{\mathrm{imp}}\approx \ln 2$ across $T$. Here $d,\delta$ are RG-invariant parameters defining the phase (see main text).
}
  \label{fig:imp-entropy}
\end{figure}

In this Letter, we determine the finite-temperature thermodynamics of a single magnetic impurity at the edge of a superconducting wire with open boundary conditions.  We note that for a system with open boundary conditions and a boundary impurity, the free energy separates into bulk, edge, and impurity parts; here we compute the impurity contribution, which is directly measurable~\cite{hartman2018direct,child2022entropy,kealhofer2025entropy}. The phase diagram consists of four distinct regimes: Kondo, YSR I, YSR II, and local moment, separated by boundary quantum phase transitions \cite{pasnoori2022rise}.  We show here that the different phases restructure the Hilbert space into different excitation towers, which manifest themselves in unusual thermodynamic properties. For example, in the YSR regimes, thermal activation of midgap bound states produces pronounced nonmonotonic entropy behavior, with overshoots above the free-moment value $\ln 2$, marking a qualitative distinction from the conventional Kondo problem characterized by a monotonic flow of impurity entropy, see Fig.~\ref{fig:imp-entropy}.

   We begin by introducing the model Hamiltonian and its Bethe Ansatz solution. The discussion then turns to the structure of the Hilbert space in the different impurity regimes, which governs the thermodynamic behavior. From this perspective, we derive and solve the thermodynamic Bethe Ansatz equations and present the resulting impurity contribution to the free energy and entropy across the different phases.

	The  superconducting wire with an impurity, studied in Ref.~\cite{pasnooriKondoSup,pasnoori2022rise,pasnoori2025emergent} is described by the Hamiltonian  
    \begin{align}
		H&=\int_{-L}^0\rm{d}x\bigl[-i\psi^\dagger_{\sigma,+}\partial_x\psi_{\sigma,+}
		+i\psi^\dagger_{\sigma,-}\partial_x\psi_{\sigma,-}\nonumber\\
		&+2g\psi^\dagger_{\sigma,+}\vec\sigma_{\sigma\rho}\psi_{\rho,+}
		\cdot\psi^\dagger_{\sigma',-}\vec\sigma_{\sigma'\rho'}\psi_{\rho',-}\nonumber\\
		&+J\delta(x)\psi^\dagger_{\sigma,-}\vec\sigma_{\sigma\sigma'}\psi_{\sigma',+}\cdot\vec S
		\bigr],
		\label{GN-ham}
	\end{align}
	where $\psi_{\sigma,\pm}(x)$ are the chiral spin$-\frac12$ fermion fields with $\pm$ denoting left/right chirality and  $\sigma=\uparrow,\downarrow$ denoting spin, $\vec\sigma=(\sigma^x,\sigma^y,\sigma^z)$ are the Pauli matrices, $\vec S$ is a spin$-\frac12$ impurity  localized at $x=0$, and summation is implied over repeated spin indices.
    
\begin{figure}
		\centering		\includegraphics[width=\linewidth]{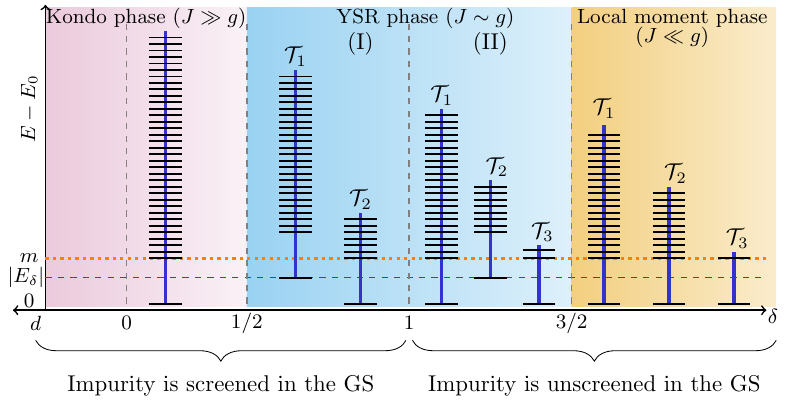}
		\caption{The Hilbert space fragments into different towers $\mathcal{T}_i$ of spin excitations in the various phases, labeled by the RG variant parameter $d$ and here $\delta=id$.}
		\label{fig:Towers}
	\end{figure}

	 The Hamiltonian is integrable, and the complete spectrum can be obtained via the Bethe Ansatz technique. The $N$ electron states are characterized by quasimomenta $k_j$  and spin rapidities $\lambda_\alpha$ that satisfy the Bethe-Ansatz equations (for derivation see Supplementary Material \cite{Supplemental}):
    \begin{eqnarray} 
    && e^{-2ik_j L}=\prod_{\upsilon=\pm}\frac{b+\upsilon \lambda_\alpha+\frac{i}{2}}{b+\upsilon \lambda_\alpha-\frac{i}{2}}, \label{cbae-cGN-Kondo}\\
&&\prod_{\upsilon=\pm} 
		\left( 
		\frac{\lambda_\alpha + \upsilon b + \frac{i}{2}}{\lambda_\alpha + \upsilon b - \frac{i}{2}} 
		\right)^N
		\left( 
		\frac{\lambda_\alpha + \upsilon d + \frac{i}{2}}{\lambda_\alpha + \upsilon d - \frac{i}{2}} 
		\right)= \nonumber \\
		&& \quad \quad \prod_{\upsilon=\pm} 
		\prod_{\substack{\beta=1 \\ \beta \ne \alpha}}^M 
		\left( 
		\frac{\lambda_\alpha + \upsilon \lambda_\beta + i}{\lambda_\alpha + \upsilon \lambda_\beta - i} 
		\right),\label{sbae-cGN-Kondo}
    \end{eqnarray} 
    where $M$ dictates the total $z-$component of the given state via $S^z=\frac{N+1}{2}-M$. It is convenient, as shown in Ref.\cite{pasnooriKondoSup,pasnoori2022rise}, to parameterize the bulk running coupling constant $b=\frac{4-g^2}{8g}$ and the boundary running coupling constant via $c=\frac{2J}{1-3J^2/4}$. While both couplings, $g$ and $J$ (and hence $b$ and $c$) are RG running, there exists a specific combination of these two running parameters $d(J,g)=\sqrt{b^2-2b/c-1}$ that is RG invariant. This RG-invariant quantity governs the impurity phases, as illustrated in Fig.~\ref{fig:Towers}.

The analysis ~\cite{pasnoori2022rise} of the coupled equations (Eq.~\eqref{cbae-cGN-Kondo} and Eq.~\eqref{sbae-cGN-Kondo}) shows that the bulk excitations have a finite gap $m = D\sinh[1/\sinh(\pi b)]$. The impurity physics, on the other hand, falls into several phases. It is governed by an RG-invariant parameter $d$, which can be real or purely imaginary, $d=i\delta$. For real $d$ or for $\delta \in [0,1/2)$, the impurity is completely screened by the gapped quasiparticles, defining the Kondo regime. When $\delta>1/2$, a YSR bound state lies inside the gap with energy $E_\delta=-m\sin(\pi\delta)$. 
For $1/2<\delta<1$, $E_\delta\in(-m,0)$ and the screened singlet is the ground state. 
At $\delta=1$, $E_\delta=0$, marking the parity-changing transition. 
For $1<\delta<3/2$, $E_\delta\in(0,m)$: the ground state is the unscreened doublet, while the screened singlet becomes an excited in-gap state. Finally, for $\delta>3/2$, the bulk superconducting correlations overwhelm the Kondo scale and the impurity remains unscreened at all energies.

However, beyond this basic classification of screening, the spectrum in each regime organizes into distinct towers of states, as shown in Fig.~\ref{fig:Towers}, whose identification is indispensable for understanding the thermodynamics, and we now turn to their structure.
In the Kondo regime, the BAE admits only conventional string solutions
\begin{equation}
    \lambda_{\alpha,j}^{(n)}=\Lambda_\alpha+\frac{i}{2} \left(n+1-2j\right),   j=1,\dots,n,\ \ \Lambda_\alpha\in\mathbb{R},
\end{equation}
which span the full Hilbert space. The ground state $|K\rangle$ is a sea of $\frac{N+1}{2}$ distinct 1-strings, and all spin excitations are spinons that arise by adding an even number of holes and higher strings, so the spectrum forms a single tower built on $|K\rangle$. For $ \tfrac{1}{2} < \delta < 1 $ (YSR-I), the solution contains, besides the usual holes and strings, a purely imaginary boundary root given by $\lambda_\delta = i\left(\delta - \frac{1}{2}\right)$.  As a result, the Hilbert space splits into two towers: $\mathcal{T}_1$ built on $|U\rangle$ (all real roots) containing $3/4$ of the total states, and $\mathcal{T}_2$ built on $|B\rangle$ (real roots plus the boundary root $\lambda_\delta$) containing $1/4$ of the total states. For $1<\delta<3/2$ (YSR-II), boundary strings are present, $\lambda_{\delta,\ell}^{(p)}=\lambda_\delta+i\ell$ with $\ell=1,\dots,p$ and $p=\lfloor\delta+\frac{1}{2}\rfloor$ (where $\lfloor\cdot\rfloor$ is the floor function), so the structure consists of three towers: $\mathcal{T}_1$ built on $|U\rangle$ (all real roots) containing $4/6$ of the total states, $\mathcal{T}_2$ built on $|B\rangle$ (real roots plus $\lambda_\delta$) containing $1/4$ of the total states, and $\mathcal{T}_3$ built on $|\tilde U\rangle$ (real roots plus $\lambda_\delta$ and the boundary strings) containing $1/12$ of the total states. For $\delta>3/2$ the same three solution classes persist and the decomposition remains threefold: $\mathcal{T}_1$ (built on the all-real solutions) containing $\frac{\lfloor 2\delta \rfloor + 2}{2(\lfloor 2\delta \rfloor+1)}$ of the states, $\mathcal{T}_2$ (built on top the state with the boundary root in addition to all real roots) containing $\frac{\lfloor 2\delta \rfloor -1}{2\lfloor 2\delta \rfloor }$ of the states, and $\mathcal{T}_3$ (built on the state with the root together with boundary strings on top of all real roots solution) containing $\frac{1}{2\lfloor 2\delta \rfloor(\lfloor 2\delta \rfloor + 1)}$ of the states. This constitutes one of our new results, providing a systematic way to construct complete sets of states within the string hypothesis across different phases.

    Having briefly presented results concerning the ground state and excitation spectrum, we now turn to the study of the impurity thermodynamics of the model. It will allow, in particular, to describe the 
    full crossover from the ultraviolet to the infrared regime in the impurity behavior in all four phases. The thermodynamics can be obtained from the thermodynamic Bethe Ansatz (TBA) equations~\cite{yang1969thermodynamics,takahashi1999thermodynamics, andrei1983solution, tsvelick1983exact, Supplemental},
	\begin{equation}
		\ln \eta_n(\lambda) = -\frac{m}{T}\cosh(\pi \lambda)\delta_{n,1}+\sum_{\upsilon=\pm 1}G\ln(1+\eta_{n+\upsilon })
		\label{TBAeqn}
	\end{equation}
	with boundary conditions $\eta_0(\lambda)=0$ and
	\begin{equation}
		\lim_{n\to \infty}\left\{[n+1]\ln(1+\eta_n)-[n]\ln(1+\eta_{n+1})\right\}=-\frac{\mu h}{T}.
		\label{tbainftybc}
	\end{equation}
    where  $h$ is the applied global magnetic field and $\mu$ is the magnetic moment. 
	Here, $[n]f(\nu)=\int \rm{d}\lambda \frac{1}{\pi} \frac{{n}/{2}}{\left(n/2\right)^2 + (\nu-\lambda)^2}f(\lambda)$ and $Gf(\lambda)=\int \rm{d}\nu \frac{1}{2\cosh\pi(\lambda-\nu)} f(\nu)$. The free energy is then given in terms of the thermodynamic functions $\eta_n(\lambda)$ as specified below.

\begin{figure}
		\centering		\includegraphics[width=\linewidth]{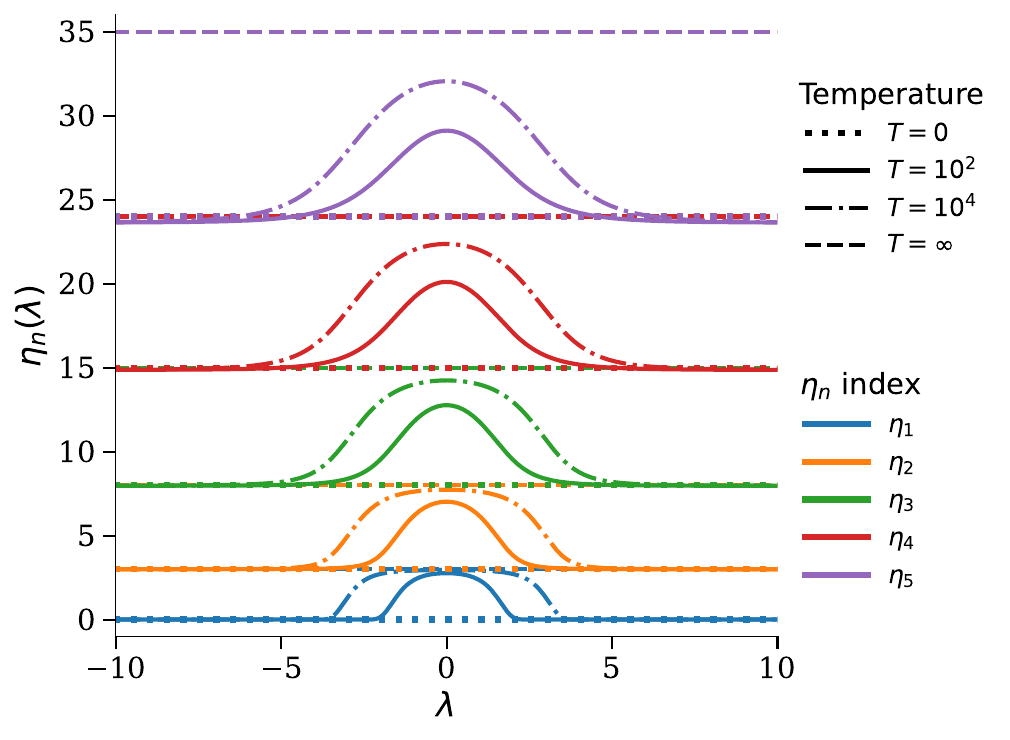}
		\caption{Numerical solutions of the TBA for $T = 100$ and $T = 1000$, with analytic limits at $T = 0$ and $T \to \infty$ for $m=1$.}
		\label{fig:eta_n(T)}
	\end{figure} 
    
    To determine the $\eta_n(\lambda)$ functions the TBA equations need to be numerically solved for each temperature $T$ due to the explicit mass scale $m$ (cf.\ Ref.\ \cite{PhysRevLett.49.497,zarand2002thermodynamics}) as shown in Fig.~\ref{fig:eta_n(T)} for some representative $T$ (see Ref.\cite{Supplemental} for details of numerical solution). However, analytic results exist in asymptotic limits ($T\to 0,\infty$), where $\eta_n(\lambda)$ becomes constant. At high temperature ($T\to\infty$), Eq.~\eqref{TBAeqn} simplifies to
	\begin{equation}
		{\eta_{n}}_{(T\to \infty)}= \frac{\sinh^2\left((n+1)\frac{\mu h}{T}\right)}{\sinh^2\frac{\mu h}{T}} - 1,\quad n=1,2,\dots,
		\label{etasTinfty}
	\end{equation}
	and at zero temperature  it reduces to
	\begin{equation}
		{\eta_{n}}_{(T\to 0)}=\frac{\sinh^2\left(n\frac{\mu h}{T}\right)}{\sinh^2\frac{\mu h}{T}} - 1,
		\quad n=1,2,\dots,
		\label{etasT0}
	\end{equation}

	As the parameter $d$ (or $\delta$) is varied, the model undergoes a boundary eigenstate phase transition, reflected in the free energy contribution due to impurity defined from the ratio of partition functions with and without impurity. For an open wire of length $L$, we write $Z_0 = e^{-\beta F_{\rm bulk}(T)} \mathfrak g(T)$, where $\mathfrak g(T)$ denotes the universal $O(1)$ boundary factor~\cite{pozsgay2010mathcal,he2024exact}. With the impurity present the partition function becomes $Z = Z_0 \mathfrak g_{\rm imp}(T)$, so that $F_{\rm imp}(T) = -\tfrac{1}{\beta}\ln(Z/Z_0) = -\tfrac{1}{\beta}\ln \mathfrak g_{\rm imp}(T)$. Since our focus is on the impurity contribution, we do not evaluate the open-edge contribution $\mathfrak g(T)$, which can be obtained following the discussion in \cite{Supplemental} (see also \cite{pozsgay2010mathcal,he2024exact}). The impurity entropy follows as $S_{\rm imp}(T) = -\partial_T F_{\rm imp}(T) = \ln \mathfrak g_{\rm imp}(T) + T \partial_T\ln \mathfrak g_{\rm imp}(T)$. At an RG fixed point, $\mathfrak g_{\rm imp}(T)$ is stationary so that $S_{\rm imp}=\ln \mathfrak g_{\rm imp}$, the Affleck--Ludwig boundary degeneracy~\cite{PhysRevLett.67.161}.

	We proceed to discuss the free energy in the various phases. In the Kondo phase—corresponding to $d \in \mathbb{R}$ or $d = i\delta$ with $\delta \in \left[0, \frac{1}{2}\right)$—all eigenstates are organized into a single tower of excitations made by adding even number of holes and bulk string solutions on top of the ground state. As a result, the impurity's free energy when $h=0$ takes the form (see \cite{Supplemental} for details)
	\begin{equation}
		F_{\rm{imp}} = -\frac{T}{4} \sum_{\upsilon = \pm} \int \frac{ \ln(1 + \eta_1) }{\cosh\left(\pi(\lambda + \upsilon d)\right)} d\lambda.
	\end{equation}

	As $\delta$ increases and enters the range $\delta \in \left(\frac{1}{2}, 1\right)$, the system transitions into the YSR regime I. In this phase, the impurity becomes screened by a boundary-bound state in the ground state. Consequently, the spectrum divides into two distinct excitation towers as discussed earlier and shown in Fig.~\ref{fig:Towers}. Accordingly, the zero-field free energy of the impurity now receives contributions from both excitation towers. As the Hilbert space splits into towers $\mathcal{T}_1$ and $\mathcal{T}_2$, the partition function can be written as $Z=\sum_{i=1}^2 e^{-\beta F_{\mathcal{T}_i}}=e^{-\beta F_{\rm bulk}(T)} \mathfrak g(T)\sum_{i=1}^2 e^{-\beta F_{\rm imp}^{(\mathcal{T}_i)}(T)}$. Defining, $\mathfrak g_{\rm imp}(T)\equiv \sum_{i=1}^2 e^{-\beta F_{\rm imp}^{(\mathcal{T}_i)}(T)}=e^{-\beta F_{\rm imp}(T)}$ gives the total impurity free energy $F_{\rm imp}(T)$. The two individual contributions (derived explicitly in Ref.~\cite{Supplemental}) are
	\begin{align}
    F_{\rm imp}^{(\mathcal{T}_1)}&=F_{\rm imp}^{(\mathcal{T}_2)}+|E_\delta|-\int \rm{d} \lambda \sum_{\upsilon=\pm}\frac{\frac{T}{4}\ln \left[1+\eta_2(\lambda)\right]}{\cosh(\pi(\lambda+i\upsilon (\delta-\frac12)))}\nonumber\\
		F_{\rm imp}^{(\mathcal{T}_2)}&=-\frac{T}{4}\int \rm{d} \lambda \sum_{\upsilon=\pm}\frac{\ln \left[1+\eta_1(\lambda)\right]}{\cosh(\pi(\lambda+i\upsilon \delta))}
		\label{freeengsT}
	\end{align}
	where $E_\delta$ is the energy of the single particle bound mode. Since impurity is screened at ground state and it is unscreened at $T\to\infty$, we expect that the impurity entropy goes from 0 at $T=0$ to $\ln 2$ at $T\to\infty$ $\forall d \cup \delta\in (0,1)$. We obtain $\eta_1$ and $\eta_2$ numerically by solving the TBA equations (Eq.~\eqref{TBAeqn}), from which we compute the impurity entropy $S_{\rm imp}(T)$. In the Kondo phase, $S_{\rm imp}(T)$ grows monotonically with $T$ (Fig.~\ref{fig:imp-entropy}) and exhibits the same critical exponents as the conventional Kondo problem (see \cite{Supplemental}). Remarkably, this monotonic behavior persists even though the $g$-theorem does not apply in a gapped bulk~\cite{PhysRevLett.67.161,Friedan2004Boundary}. By contrast, in the YSR I phase $S_{\rm imp}(T)$ shows a non-monotonic hump at intermediate $T$, even exceeding the free-moment value of $\ln 2$ for larger $\delta$ due to the presence of the midgap YSR states, as shown in Fig.~\ref{fig:imp-entropy}.

	The overshoot of the impurity entropy above $\ln2$ at intermediate temperatures follows directly from writing
	$S_{\rm imp}^{(\mathcal{T}_i)}(T) = -\frac{\rm{d}}{\rm{d}T}F_{\rm imp}^{(\mathcal{T}_i)}$
	and using Eqs.~\eqref{etasT0}–\eqref{etasTinfty} to obtain
	$S_{\rm imp}^{(\mathcal{T}_2)}(0)=0,\quad
	S_{\rm imp}^{(\mathcal{T}_2)}(\infty)=-\ln2,\quad
	S_{\rm imp}^{(\mathcal{T}_1)}(0)=\ln2,\quad
	S_{\rm imp}^{(\mathcal{T}_1)}(\infty)=\ln\frac32.$
	One then finds the total impurity entropy from the total impurity free energy $F_{\rm imp}$, which satisfies
$e^{\beta F_{\rm imp}}=e^{\beta F_{\rm imp}^{(\mathcal{T}_1)}}+e^{\beta F_{\rm imp}^{(\mathcal{T}_2)}}$
 such that
$S_{\rm imp}=-\frac{\rm d}{\rm dT}F_{\rm imp}$ gives
\begin{equation}
    S_{\rm imp}(T)
    = \ln\bigl(\mathscr{Z}\bigr)
    + \frac{|E_\delta|}{T} 
    \frac{e^{S_{\rm imp}^{(\mathcal{T}_1)} - \beta |E_\delta|}}
    {\mathscr{Z}},
    \label{stot-t1-t2}
\end{equation}
	with $\mathscr Z=e^{S_{\rm imp}^{(\mathcal{T}_1)}-\beta|E_\delta|}+e^{S_{\rm imp}^{(\mathcal{T}_2)}}$ as the tower $\mathcal{T}_1$ is lifted in energy by $|E_\delta|$. The first, weighted “ensemble” average (mixing) term increases smoothly from 0 to $\ln2$, while the second, “activation” term is vanishing for $T\ll|E_\delta|$ and $T\gg|E_\delta|$, but it peaks at $T\sim|E_\delta|$ when the impurity‐induced YSR bound state becomes thermally accessible and it briefly drives $S_{\rm imp}$ above $\ln2$ (inset of Fig.~\ref{fig:imp-entropy}, $\delta=0.9$). More precisely,  the impurity plays a twofold role: at low temperature, it is fully adsorbed (hybridized) into the bulk, yielding vanishing entropy, while at high temperature it is asymptotically free, contributing $\ln 2$ entropy, mirroring the monotonic crossover known from the conventional Kondo effect. Additionally, the impurity induces a midgap bound state absent in the clean system, whose thermal population produces a transient entropy bump near its energy scale, causing the total impurity entropy to exhibit non-monotonic behavior. We also demonstrate this overshoot numerically in a related lattice model (in \cite{Supplemental}). 
	
	   At $\delta=1$ the system undergoes a parity‐changing boundary quantum phase transition: for $\delta<1$ the screened singlet is the ground state and the unscreened doublet lies at midgap, whereas for $1<\delta<3/2$ (YSR II) the roles reverse, with the unscreened doublet forming the ground state and the screened singlet appearing as an excited midgap state. For all $\delta>1$ the eigenstates arrange into three distinct towers of excitation as shown in Fig.~\ref{fig:Towers}. In the YSR II regime the tower $\mathcal{T}_2$ built from the fundamental boundary root $\lambda_\delta=i(\delta-\tfrac{1}{2})$ is lifted in energy, since this root carries a positive level $E_\delta=m>-m\sin(\pi\delta)>0$, while the higher–order boundary string $\lambda_{\delta,\ell}^{(p=1)}$ has energy $E_{\delta,p=1}=m\sin(\pi\delta)$. In the local moment regime $\delta>\frac{3}{2}$, energies of both fundamental boundary root and higher-order boundary strings vanish. The impurity contribution to the zero–field free energy then receives separate parts from $\mathcal{T}_1$, $\mathcal{T}_2$, and $\mathcal{T}_3$ in $\delta>1$ regime, where


\begin{align}
F_{\rm imp}^{(\mathcal{T}_1)} &= - \frac{T}{4} \sum_{\upsilon = \pm} \int \mathrm{d}\lambda 
\left[
\frac{\ln \left(1 + \eta_{\lfloor 2\delta \rfloor +1}(\lambda)\right)}
     {\cosh \left(\pi\left(\lambda + \frac{i\upsilon}{2}(2\delta - \lfloor 2\delta \rfloor)\right)\right)}
\right. \nonumber\\
&\hspace{4.5em}\left.
- \frac{\ln \left(1 + \eta_{\lfloor 2\delta \rfloor}(\lambda)\right)}
     {\cosh \left(\pi\left(\lambda + \frac{i\upsilon}{2}(\lfloor 2\delta \rfloor+1 - 2\delta)\right)\right)}
\right], \nonumber\\
F_{\rm imp}^{(\mathcal{T}_2)} &= \frac{T}{4} \sum_{\upsilon = \pm} \int \mathrm{d}\lambda 
\left[
\frac{\ln \left(1 + \eta_{\lfloor 2\delta \rfloor - 1}(\lambda)\right)}
     {\cosh \left(\pi\left(\lambda + \frac{i\upsilon}{2}(2\delta - \lfloor 2\delta \rfloor)\right)\right)}
\right. \nonumber\\
&\left.
- \frac{\ln \left(1 + \eta_{\lfloor 2\delta \rfloor - 2}(\lambda)\right)}
     {\cosh \left(\pi\left(\lambda + \frac{i\upsilon}{2}(\lfloor 2\delta \rfloor+1 - 2\delta)\right)\right)}
\right]+E_\delta, \nonumber\\
F_{\rm imp}^{(\mathcal{T}_3)} &= \frac{T}{4} \sum_{\upsilon = \pm} \int \mathrm{d}\lambda 
\left[
\frac{\ln \left(1 + \eta_{\lfloor 2\delta \rfloor}(\lambda)\right)}
     {\cosh \left(\pi\left(\lambda + \frac{i\upsilon}{2}(\lfloor 2\delta \rfloor+1 - 2\delta)\right)\right)}
\right. \nonumber\\
&\hspace{4.5em}\left.
+ \frac{\ln \left(1 + \eta_{\lfloor 2\delta \rfloor - 1}(\lambda)\right)}
     {\cosh \left(\pi\left(\lambda + \frac{i\upsilon}{2}(2\delta - \lfloor 2\delta \rfloor)\right)\right)}
\right]. \nonumber
\end{align}

	
	In the YSR II regime $(1<\delta<\frac{3}{2})$, with $E_{\delta}=m\sin\left(\pi\delta\right)>0$, using Eq.~\eqref{etasT0} and Eq.~\eqref{etasTinfty} one finds $S_{\rm imp}^{(\mathcal{T}_1)}(0)=\ln\left(\frac{3}{2}\right)$, $S_{\rm imp}^{(\mathcal{T}_2)}(0)=0$, $S_{\rm imp}^{(\mathcal{T}_3)}(0)=-\ln 2$, and $S_{\rm imp}^{(\mathcal{T}_1)}(\infty)=\ln\left(\frac{4}{3}\right)$, $S_{\rm imp}^{(\mathcal{T}_2)}(\infty)=-\ln 2$, $S_{\rm imp}^{(\mathcal{T}_3)}(\infty)=-\ln 6$. Since $E_{\delta}>0$ lifts the $\mathcal{T}_2$ tower, at $T=0$ only $\mathcal{T}_1$ and $\mathcal{T}_3$ contribute, giving $S_{\rm imp}(0)=\ln\left(e^{S_{\rm imp}^{(\mathcal{T}_1)}(0)}+e^{S_{\rm imp}^{(\mathcal{T}_3)}(0)}\right)=\ln 2$. In the high–temperature limit, all three towers contribute, so $S_{\rm imp}(\infty)=\ln\left(e^{S_{\rm imp}^{(\mathcal{T}_1)}(\infty)}+e^{S_{\rm imp}^{(\mathcal{T}_2)}(\infty)}+e^{S_{\rm imp}^{(\mathcal{T}_3)}(\infty)}\right)=\ln 2$. As shown in Fig.~\ref{fig:imp-entropy}, the impurity entropy exceeds $\ln 2$ at intermediate temperatures $T\sim E_{\delta}$ when the midgap $\mathcal{T}_2$ states become thermally accessible.


As $\delta>\frac{3}{2}$ the energies of all boundary string solutions $\lambda_{\delta}$ and $\lambda_{\delta,\ell}^{(p)}$ vanish in the thermodynamic limit, so all three towers are available even at low temperatures. Using Eq.~\eqref{etasT0} and Eq.~\eqref{etasTinfty} one finds $S_{\rm imp}^{(\mathcal{T}_1)}(0)=\ln\left(\frac{\left\lfloor 2\delta \right\rfloor+1}{\left\lfloor 2\delta \right\rfloor}\right)$, $S_{\rm imp}^{(\mathcal{T}_2)}(0)=\ln\left(\frac{\left\lfloor 2\delta \right\rfloor-2}{\left\lfloor 2\delta \right\rfloor-1}\right)$, $S_{\rm imp}^{(\mathcal{T}_3)}(0)=\ln\left(\frac{1}{\left(\left\lfloor 2\delta \right\rfloor-1\right)\left\lfloor 2\delta \right\rfloor}\right)$, and $S_{\rm imp}^{(\mathcal{T}_1)}(\infty)=\ln\left(\frac{\left\lfloor 2\delta \right\rfloor+2}{\left\lfloor 2\delta \right\rfloor+1}\right)$, $S_{\rm imp}^{(\mathcal{T}_2)}(\infty)=\ln\left(\frac{\left\lfloor 2\delta \right\rfloor-1}{\left\lfloor 2\delta \right\rfloor}\right)$, $S_{\rm imp}^{(\mathcal{T}_3)}(\infty)=\ln\left(\frac{1}{\left(\left\lfloor 2\delta \right\rfloor\right)\left(\left\lfloor 2\delta \right\rfloor+1\right)}\right)$. Thus $S_{\rm imp}(0)=\ln 2=S_{\rm imp}(\infty)$, and, as shown in Fig.~\ref{fig:imp-entropy} for representative values $\delta=1.75$ and $\delta=2.35$, the impurity entropy can exceed its free-moment value $\ln 2$ at intermediate temperatures when these towers become thermally accessible.

{\it In sum:} We presented an exact thermodynamic classification of a magnetic impurity at the edge of a gapped 1D host, where the RG-invariant parameter $\delta$ drives boundary phase transitions that reorganize the Hilbert space into distinct excitation towers. In the Kondo phase, the impurity entropy varies monotonically with temperature as in the conventional Kondo problem, whereas midgap Yu-Shiba-Rusinov states produce overshoots above $\ln 2$, demonstrating the role of discrete mid-gap boundary excitations in gapped systems in intermediate temperatures. Deep in the local moment phase, the entropy remains close to $\ln 2$ with only small, vanishing (as $\delta\to\infty$) bumps at intermediate temperatures. These results show that in a gapped 1D superconductor, a single impurity generates a richer phase diagram than in a gapless Fermi liquid, where only the Kondo phase exists; while the Kondo phase here retains the same critical exponents and monotonic impurity entropy flow as in the gapless case, additional phases with distinct thermodynamic properties also emerge.

Beyond this specific setting, our work develops a general TBA framework for handling impurities or generic defects that generate boundary-bound modes, where the associated boundary roots fragment the Hilbert space into distinct excitation towers. This provides a unifying perspective on boundary phase transitions and their thermodynamic signatures as well as their experimental manifestations~\cite{hartman2018direct,child2022entropy,kealhofer2025entropy} across a wide class of impurity~\cite{TBA-XXX-paper} and defect problems.

An important open question is whether the Hilbert-space fragmentation that drives the impurity thermodynamics also leaves fingerprints in quantum dynamics and transport, constraining relaxation and non-equilibrium steady states. Equally intriguing is whether similar Hilbert space fragmentation arises in higher-dimensional or unconventional superconductors, where most studies rely on semiclassical approaches that may have overlooked such effects.

{\it Acknowledgments:} We thank P.~Pasnoori, C.~Rylands, and P.~Azaria for collaboration in related work and useful discussions, and Y.~Tang and A.~Tsvelik for helpful discussions.

\bibliography{ref}

	\widetext
	\newpage
	\begin{center}
		\textbf{\large Supplementary Materials for `Thermodynamics in a split Hilbert space:  Quantum impurity at the edge of a one-dimensional superconductor'}\\
		\vspace{0.4cm}
		Pradip Kattel,$^1$ Abay Zhakenov,$^1$  and Natan Andrei$^1$\\
		\vspace{0.2cm}
		\small{$^1$\textit{Department of Physics and Astronomy, Center for Materials Theory, Rutgers University, Piscataway, New Jersey 08854, USA}}
	\end{center}
	
	\renewcommand{\thesection}{S.\arabic{section}}

	\setcounter{equation}{0} 
	\setcounter{page}{1}
	
	\renewcommand{\theequation}{S.\arabic{equation}}
	
	In this Supplementary Material, we provide additional details supporting the results presented in the main text. We first review the Bethe Ansatz solution of the model, including the structure of the excitation towers and the associated boundary quantum phase transitions. We then present the full derivation of the thermodynamic Bethe Ansatz (TBA) equations in each phase, together with the expressions for the free energy contribution due to impurity. Numerical methods for solving the truncated TBA equations are described, along with convergence checks. We then briefly discuss the solution of the problem with a fixed number of particles. Finally, we present exact diagonalization results for a related lattice model, which illustrate and confirm the analytical predictions.

	\section{Towers of Excitations and boundary quantum phase transition}\label{sec:towerscGNK}
	The Hamiltonian considered in the main text
    \begin{align}
		H&=\int_{-L}^0\rm{d}x\bigl[-i\psi^\dagger_{\sigma,+}\partial_x\psi_{\sigma,+}
		+i\psi^\dagger_{\sigma,-}\partial_x\psi_{\sigma,-}+2g\psi^\dagger_{\sigma,+}\vec\sigma_{\sigma\rho}\psi_{\rho,+}
		\cdot\psi^\dagger_{\sigma',-}\vec\sigma_{\sigma'\rho'}\psi_{\rho',-}+J\delta(x)\psi^\dagger_{\sigma,-}\vec\sigma_{\sigma\sigma'}\psi_{\sigma',+}\cdot\vec S
		\bigr],
	\end{align}
    is integrable for arbitrary bulk and boundary couplings as shown in Ref.~\cite{pasnoori2022rise}. The complete spectrum for arbitrary $J>0$ and $g>0$ is obtained from the Bethe Ansatz equations (BAE)
	\begin{equation}
		e^{-2ik_j L}=\prod_{\upsilon=\pm}\frac{b+\upsilon \lambda_\alpha+\frac{i}{2}}{b+\upsilon \lambda_\alpha-\frac{i}{2}}
		\label{chargeBAE-GN}
	\end{equation}
	where the rapidities $\lambda_\alpha$ for fixed particle sector $N$ satisfy coupled algebraic equations of the form
	\begin{align}
		& \prod_{\upsilon=\pm} 
		\left( 
		\frac{\lambda_\alpha + \upsilon b + \frac{i}{2}}{\lambda_\alpha + \upsilon b - \frac{i}{2}} 
		\right)^N
		\left( 
		\frac{\lambda_\alpha + \upsilon d + \frac{i}{2}}{\lambda_\alpha + \upsilon d - \frac{i}{2}} 
		\right)=\prod_{\upsilon=\pm} 
		\prod_{\substack{\beta=1 \\ \beta \ne \alpha}}^M 
		\left( 
		\frac{\lambda_\alpha + \upsilon \lambda_\beta + i}{\lambda_\alpha + \upsilon \lambda_\beta - i} 
		\right).\label{spinBAE_GN}
	\end{align}
	Here, $M$ dictates the total $z-$component of the given state via $S^z=\frac{N+1}{2}-M$.
	
	Under the string hypothesis, one assumes that all the solutions of Eq.\eqref{spinBAE_GN} are of the form
	\begin{equation}\label{strings}
		\left\{\lambda_{\alpha,j}^{(n)}\right\}_{j=1}^n
		=
		\left\{\Lambda_\alpha + \frac{i}{2}(n+1 - 2j)
		\middle|  j = 1,\dots,n\right\},
	\end{equation}
	where $\Lambda_\alpha \in \mathbb{R}$ is the center of each strings of length $n$. Apart from these usual string solutions, there exist unique boundary string solutions of the form
	\begin{equation}\label{boundary-string}
		\lambda_{\delta,\ell}^{(p)}=\lambda_\delta+i\ell ~\quad\text{for}\quad \ell=1,\cdots,p,
	\end{equation}
	where $p=$ the fundamental boundary string $\lambda_\delta$ is a purely imaginary solution of the Bethe Ansatz equations Eq.\eqref{spinBAE_GN} of the form $i\left(\frac{1}{2}-\delta\right)$ which exists only when $\delta>\frac{1}{2}$ and $m=\lceil\delta-\frac{1}{2}\rceil$. While the full solution was given in Ref.~\cite{pasnoori2022rise}, we briefly review the solution pertaining to the towers of excitations and boundary phase transitions essential for computing the impurity contribution to the free energy.
	
	\subsection{The Kondo phase}
	When $d\in \mathbb{R}$ or $d=i\delta$ and $\delta\in\left(0,\frac{1}{2}\right)$, the Kondo coupling $J$ is stronger than the ulk superconducting coupling $g$ and hence according to the beta function $\beta(J)=-\frac{2}{\pi}J(J-g)$ (see Eq.\eqref{betaJeqn} and Ref.\cite{pasnooriKondoSup}), the Kondo coupling grows as the energy scale is lowered and hence despite the gap in the spectrum, the impurity is completely screened by manybody Kondo cloud.  Taking $\ln$ on both sides of Eq.\eqref{chargeBAE-GN} and Eq.\eqref{spinBAE_GN}, we write the Bethe Ansatz equations in Logarithmic form as
	\begin{equation}
		k_j=\frac{\pi n_j}{L}+\frac{1}{L}\sum_{\upsilon}\arctan\left(2(b+\upsilon\lambda) \right)
	\end{equation}
	where $n_j$ are the charge quantum numbers, and the energy is given by
	\begin{equation}
		E=\sum_j k_j=\frac{\pi}{L}\sum_j n_j+D\sum_{\upsilon=\pm}\arctan(2(b+\upsilon \lambda_\alpha)),\label{engEqnGNK}
	\end{equation}
	and the relation for spin rapidities becomes
	\begin{equation}
		\sum_{\upsilon=\pm} \left[N\arctan(2(\lambda_\alpha+\upsilon b))+\arctan(2(\lambda_\alpha+\upsilon d))\right]=\sum_{\upsilon=\pm} \sum_{\beta\neq \alpha}\arctan(\lambda_\alpha+\upsilon\lambda_\beta)+\pi I_j.
		\label{logBAE-spin-GN}
	\end{equation}
	Here, $I_j\in \mathbb{Z}$ are spin quantum numbers. 
	Taking the derivative of Eq.\eqref{logBAE-spin-GN}, we obtain an integral equation for the root density in the ground state as
	\begin{equation}
		2\rho_{\ket{K}}= \frac{1}{\pi}\sum_{\upsilon=\pm}\left( \frac{2N}{4 (b \upsilon +\lambda)^2+1}+\frac{2}{4 (d \upsilon +\lambda )^2+1}\right)+\frac{1}{\pi}\frac{2}{4\lambda^2+1}-\frac{1}{\pi}\int \mathrm{d}\lambda' \rho_{\ket{K}}\frac{1}{(\lambda-\lambda')^2+1}-\delta(\lambda),
	\end{equation}
	where the $\delta(\lambda)$ is included to remove a root at $\lambda=0$ which trivially  solved the Bethe Ansatz equations Eq.\eqref{spinBAE_GN} but leads to a nonnormalizable Bethe vector. The above integral equation can immediately be solved in Fourier space to obtain the root density
	\begin{equation}
		\tilde\rho_{\ket{K}}(\omega)=\frac{1}{4} \text{sech}\left(\frac{\omega  }{2}\right)
		\left(2 N \cos (b \omega )+2 \cos
		(d \omega )+1-e^{\frac{| \omega | }{2}}\right).
	\end{equation}
	
	The total number of Bethe roots in the ground state is $M=\int\rho_{\ket K}(\lambda)\mathrm{d}\lambda=\tilde\rho_{\ket K}(0)=\frac{N+1	}{2}$ such that the spin if the ground state is $S^z=\frac{N+1	}{2}-M=0$. This suggests that the ground state is a many-body singlet where the impurity is completely screened by the many-body Kondo screening cloud. Moreover, the fermionic parity operator $\mathcal{P}$ is defined by $\mathcal{P} = (-1)^N$, where $N$ is the total fermion number. This can equivalently be written in terms of the spin $z$-component as $\mathcal{P} = (-1)^{2 S^z - 1}$. Therefore, for the state $\ket{K}$, the fermionic parity is $\mathcal{P} \ket{K} = - \ket{K}$.

The gapless charge excitations are constructed by varying the quantum number $n_j$, whereas all the spin excitation on top of the state $\ket{K}$ is constructed by adding an even number of holes and the bulk string solution Eq.\eqref{strings}. Thus, all the excitations are sorted in a single tower of excited states. For example, the simplest spin excitation is constructed by adding two holes on top of $\ket{K}$, which results in the change in root density of the form
\begin{equation}
	\Delta\tilde\rho_{\theta_1,\theta_2}=-\frac{1}{2}\text{sech}\left(\frac{\omega}{2}\right)(\cos(\omega\theta_1+\cos(\omega\theta_2))e^{\frac{|\omega|}{2}},
\end{equation}
such that the energy of this excitation using Eq.\eqref{engEqnGNK} is
\begin{equation}
E_{\theta_1,\theta_2} = D \left[ \arctan\left( \frac{\cosh(\pi \theta_1)}{\sinh(\pi b)} \right) + \arctan\left( \frac{\cosh(\pi \theta_2)}{\sinh(\pi b)} \right) \right].
\end{equation}
We obtain the universal answer by taking the UV cutoff $D\to\infty$ and the coupling $b\to\infty$, in this double limit, we hold $m=D\arctan\left(\frac{1}{\sinh(\pi b)}\right)$ fixed and obtain the relativistic dispersion for the spinons i.e.
\begin{equation}
	E_{\theta_1,\theta_2}=m\cosh(\pi \theta_1)+m\cosh(\pi \theta_2),
\end{equation}
which shows that the spin excitations are gapped.  Likewise, all other spin excitations are constructed by adding an even number of holes and bulk string solutions. We shall see that this construction changes in other phases. 

\subsection{The YSR phase I}
When the RG invariant parameter $\delta$ takes a value in the range $\frac{1}{2}<\delta<1$, then the model is in the YSR phase. Here the bulk and boundary coupling strengths are comparable $g\sim J$ which leads to the formation of mid-gap YSR state. 

The Bethe Ansatz equations in this phase take the following form
\begin{align}
	&\left( 
	\frac{\lambda_\alpha + i \left(\delta + \frac{1}{2}\right)}{\lambda_\alpha -i \left(\delta + \frac{1}{2}\right)} 
	\right)\left( 
	\frac{\lambda_\alpha -i \left(\delta - \frac{1}{2}\right)}{\lambda_\alpha + i \left(\delta- \frac{1}{2}\right)} 
	\right) \prod_{\upsilon=\pm} 
	\left( 
	\frac{\lambda_\alpha + \upsilon b + \frac{i}{2}}{\lambda_\alpha + \upsilon b - \frac{i}{2}} 
	\right)^N
	=\prod_{\upsilon=\pm} 
	\prod_{\substack{\beta=1 \\ \beta \ne \alpha}}^M 
	\left( 
	\frac{\lambda_\alpha + \upsilon \lambda_\beta + i}{\lambda_\alpha + \upsilon \lambda_\beta - i} 
	\right).\label{spinBAE_GN-U}
\end{align}

Following the same procedure as above, we obtain the root density of the form
\begin{equation}
	\tilde\rho_{\ket{U}}(\omega)=\frac{1}{4} \text{sech}\left(\frac{\omega  }{2}\right)
	\left(2 N \cos (b \omega )+\left(1-e^{| \omega | }\right) e^{-\delta  | \omega | }+1-e^{\frac{| \omega | }{2}}\right),
\end{equation}
such that the total number of roots in this state $\ket{U}$ is $M_{\ket U}=\frac{N}{2}$ and hence the spin of this state is $S^z_{\ket{U}}=\frac{N+1}{2}-M_{\ket{U}}=\frac{1}{2}$ which shows that this is a doublet state where impurity is unscreened. The fermionic parity of this state is $\mathcal{P}=1$.

As mentioned earlier, in this regime, there also exists a unique purely imaginary solution of the Bethe Ansatz equation $\lambda_\delta$, adding this solution, the Bethe Ansatz equation becomes

\begin{align}
	&\left( 
	\frac{\lambda_\alpha - i \left(\frac{3}{2}-\delta\right)}{\lambda_\alpha +i \left(\frac{3}{2}-\delta\right)} 
	\right)\left( 
	\frac{\lambda_\alpha -i \left(\delta - \frac{1}{2}\right)}{\lambda_\alpha + i \left(\delta- \frac{1}{2}\right)} 
	\right) \prod_{\upsilon=\pm} 
	\left( 
	\frac{\lambda_\alpha + \upsilon b + \frac{i}{2}}{\lambda_\alpha + \upsilon b - \frac{i}{2}} 
	\right)^N
	=\prod_{\upsilon=\pm} 
	\prod_{\substack{\beta=1 \\ \beta \ne \alpha}}^M 
	\left( 
	\frac{\lambda_\alpha + \upsilon \lambda_\beta + i}{\lambda_\alpha + \upsilon \lambda_\beta - i} 
	\right).\label{spinBAE_GN-B}
\end{align}

Following the same procedure as above, we obtain the root density of the form
\begin{equation}
	\tilde\rho_{\ket{B}}(\omega)=\frac{1	}{4}\text{sech}\left(\frac{\omega}{2}\right)\left(2N\cosh(b\omega) -2 \cosh ((\delta -1) \omega )+1-e^{\frac{|\omega|}{2}}\right),
\end{equation}
such that the total number of roots including one purely imaginary root $\lambda_\delta$ is $M_{\ket B}=1+\int\rho(\lambda)\mathring{d}\lambda=\frac{N+1}{2}$ such that the spin of this state is $S^z_{\ket B}=\frac{N+1}{2}-M_{\ket{B}}=0$. This shows that the impurity is screened by the single particle bound mode in this state. The fermionic parity of this state is $\mathcal{P}=-1$.

Notice that the root density contribution due to the boundary string root is $\Delta\tilde\rho_\delta=-\frac{1}{4}\text{sech}\left(\frac{\omega}{2}\right)(e^{(\delta -1) | \omega | }+e^{-\delta  | \omega | })$. The energy of this state, including its bare energy, is
\begin{equation}
    E_\delta=-m\sin(\pi\delta)<0.
\end{equation}
Since, $E_{\ket{B}}-E_{\ket{U}}=E_\delta<0$, state $\ket B$ is lower energy than $\ket{U}$.

The full excitation spectrum naturally organizes into two distinct towers. Each tower is generated by adding an even number of holes and bulk string excitations on top of one of the two fundamental reference states, denoted by $\ket{U}$ and $\ket{B}$, respectively. Thus, these two base states serve as the roots for the corresponding excitation hierarchies, defining two separate and well-structured towers of excitations.

\subsection{The YSR phase II}\label{sec:YSRII-gs}
When $1 < \delta < \frac{3}{2}$, the mid-gap state persists. Both the odd fermionic parity state $\ket{B}$ and the even fermionic parity state $\ket{U}$, constructed above, remain valid eigenstates in this regime. However, now the energy difference satisfies
\begin{equation}
E_\delta = E_{\ket{B}} - E_{\ket{U}} = -m \sin(\pi \delta) > 0,
\end{equation}
implying that the doublet state $\ket{U}$, where the impurity is unscreened, is the ground state, whereas the singlet state $\ket{B}$, corresponding to the screened impurity, becomes the mid-gap state. Consequently, the model exhibits a first-order parity-changing phase transition at $\delta = 1$, such that for $\delta < 1$ the impurity is screened in the ground state, while for $\delta > 1$ the impurity remains unscreened in the ground state.

As before, two towers of excitations can be constructed by adding an even number of holes and bulk string solutions on top of these states. However, counting the states generated by these two towers shows that their total dimension does not sum to the full Hilbert space dimension $2 \times 2^N$. This indicates the necessity of an additional tower of states. It turns out that this missing tower can be constructed by including a higher-order boundary string solution, which, combined with appropriate holes and bulk strings, completes the full spectrum.

Apart from the fundamental boundary string $\lambda_\delta$, a higher order boundary string $\lambda_{\delta,1^{(1)}}=\lambda_\delta+i$ also exists in this phase. Adding this solution on top of the fundamental boundary string $\lambda_\delta$, we obtain the Bethe Ansatz equation of the form
\begin{align}
	&\left( 
	\frac{\lambda_\alpha - i \left(\frac{5}{2}-\delta\right)}{\lambda_\alpha +i \left(\frac{5}{2}-\delta\right)} 
	\right)\left( 
	\frac{\lambda_\alpha - i \left(\frac{3}{2}-\delta\right)}{\lambda_\alpha +i \left(\frac{3}{2}-\delta\right)} 
	\right)\left( 
	\frac{\lambda_\alpha -i \left(\delta - \frac{1}{2}\right)}{\lambda_\alpha + i \left(\delta- \frac{1}{2}\right)} 
	\right)^2 \prod_{\upsilon=\pm} 
	\left( 
	\frac{\lambda_\alpha + \upsilon b + \frac{i}{2}}{\lambda_\alpha + \upsilon b - \frac{i}{2}} 
	\right)^N
	=\prod_{\upsilon=\pm} 
	\prod_{\substack{\beta=1 \\ \beta \ne \alpha}}^M 
	\left( 
	\frac{\lambda_\alpha + \upsilon \lambda_\beta + i}{\lambda_\alpha + \upsilon \lambda_\beta - i} 
	\right).\label{spinBAE_GN-UU}
\end{align}

Using the same procedure mentioned above, we obtain the root density of this state as
\begin{align}
    \tilde\rho_{\ket{\tilde U}}=\frac{1}{4} \text{sech}\left(\frac{\omega }{4}\right)
   \left(2 N \cosh (b \omega
   )-e^{\left(\delta -\frac{5}{2}\right) | \omega |
   }-e^{\left(\delta -\frac{3}{2}\right) | \omega | }-2
   e^{-\left(\left(\delta -\frac{1}{2}\right) | \omega |
   \right)}-e^{\frac{| \omega | }{2}}+1\right),
\end{align}
such that the total number of roots, including the 2 complex roots, is $M_{\ket{\tilde U}}=2+\int\mathrm{d}\lambda\rho_{\ket{\tilde U}}(\lambda)=\frac{N+2}{2}$. Since $M_{\ket{\tilde U}}>\frac{N+1}{2}$, it is not a valid state, but we can construct valid excited states by adding an even number of holes and bulk string solutions on top of this state. Thus, there exist three fundamental base states, denoted by $\ket{B}$, $\ket{U}$, and $\ket{\tilde{U}}$, upon which one can construct three distinct towers of expectation values by adding an even number of holes and bulk string solutions.

\subsection{The unscreened phase}
When $\delta>\frac{3}{2}$, the bulk superconducting coupling strength $g$ is stronger than the boundary coupling $J$, and hence when the energy scale is lowered, $J$ flows to zero according to the beta function $\beta(J)=-\frac{2}{\pi}J(J-g)$. Thus, in this phase, the impurity is unscreened. 

The state $\ket{U}$, constructed above, is the ground state in this phase. The boundary string solution $\lambda_\delta$, which also exists in this phase, has vanishing energy in the thermodynamic limit. Consequently, in addition to the tower of excitations constructed on top of $\ket{U}$, there exist two additional towers of states: one constructed by adding the fundamental boundary string $\lambda_\delta$ along with an appropriate number of holes and bulk strings, and another constructed by including both $\lambda_\delta$ and a higher-order boundary string of length $p = \lceil \delta - \tfrac{1}{2} \rceil$, together with suitable holes and bulk strings, to generate the corresponding tower of excitations.

\bigskip 

After a brief discussion of the solutions pertaining to the low-lying excitations and the associated towers of states, we pause to emphasize that the change in the structure of the Hilbert space tower, as described above, is a physically meaningful phenomenon. This change manifests in all physical quantities related to the impurity. Over recent years, the effects of such transitions have been extensively studied in the thermodynamics and dynamics of spin chain models \cite{kattel2024kondo,kattel2025edge,tang2025quantum}. We shall now proceed to generalize it to the case of continuum field theory. 

We now proceed to derive the thermodynamic Bethe Ansatz (TBA) equations for the model and explicitly compute the impurity free energy.

\section{Thermodynamic Bethe Ansatz}\label{sec:TBA:GN}

We shall derive the TBA equations of the model in explicit detail within the Kondo phase and briefly outline how the construction extends to the remaining three phases.

\subsection{The Kondo Phase}
Plugging in the string solution Eq.\eqref{strings} in Bethe Ansatz equations Eq.\eqref{spinBAE_GN} upon taking $\ln$ on both sides become
\begin{equation}
  \Theta_{n}(\lambda^{(n)}_\gamma)+\sum_\upsilon  \Theta_{n}(\lambda^{(n)}_\gamma+\upsilon  d)+N \Theta_{n}(\lambda^{(n)}_\gamma+\upsilon  b)=\sum_{m,\beta}\Theta_{n,m}(\lambda^{(n)}_\gamma+\upsilon \lambda^{(m)}_\beta)-2\pi I_\gamma^{(n)},
\end{equation}

where
\begin{equation}
    \Theta_n(x)=-2 \tan ^{-1}\left(\frac{2 x}{n}\right),
\end{equation}

and

\begin{equation}
    \Theta_{mn}(x) = \begin{cases} 
\Theta_{|n-m|}(x) + 2\Theta_{|n-m|+2}(x) + \cdots + 2\Theta_{n+m-2}(x) + \Theta_{n+m}(x), & n \neq m \\ 
2\Theta_2(x) + \cdots + 2\Theta_{2n-2}(x) + \Theta_{2n}(x), & n = m .
\end{cases}
\end{equation}

The counting function 
\begin{equation}
    \begin{aligned}
        \nu_n(\lambda) = \frac{1}{2\pi}\left[\Theta_{n}(\lambda^{(n)}_\gamma)+\sum_\upsilon  \Theta_{n}(\lambda^{(n)}_\gamma+\upsilon  d)+N\Theta_{n}(\lambda^{(n)}_\gamma+\upsilon  b)-\sum_{m,\beta}\Theta_{n,m}(\lambda^{(n)}_\gamma+\upsilon \lambda^{(m)}_\beta) \right]
    \end{aligned}
\end{equation}
is such that it gives the integers $I^{(n)}_\gamma$ for corresponding roots $\lambda^{(n)}_\gamma$ \textit{i.e.} $\nu_n(\lambda^{(n)}_\gamma)=I^{(n)}_\gamma$ and gives skipped integers $I^{(n),h}_\gamma$ are the positions of holes \textit{i.e.} $\nu_n(\lambda^{(n),h}_\gamma)=I^{(n),h}_\gamma$.          
The derivative of the counting function in the thermodynamic limit gives the density of $n$-strings $\sigma_n(\mu)$ and holes $\sigma_n^h(\mu)$
\begin{equation}
    \frac{\mathrm d \nu_n}{\mathrm d \mu} = \sigma_n(\mu) + \sigma_n^h(\mu).
\end{equation}

Combining the last two expressions gives
\begin{equation}\label{sigma_n^h}
    \sigma_n^h(\mu) = f_n(\mu) - \sum_{m=1}^{\infty}A_{nm} \sigma_m(\mu).
\end{equation}
The notations used above are
\begin{equation}
    \begin{aligned}
        f_n(\mu)&=+N K_{n}(\mu- b)+N K_{n}(\mu+ b) +K_{n}(\mu+ d) +K_{n}(\mu- d)+K_{n}(\mu)\\
        A_{nm}&=\left[|n-m|\right]+2\left[|n-m|+2\right]+\ldots+2\left[n+m-2\right]+\left[n+m\right],
    \end{aligned}
\end{equation}
with $K_n(\mu)$ defined as
\begin{equation}
    K_n(\mu) \equiv -\frac{1}{2\pi} \frac{\mathrm{d}\Theta_n}{\mathrm{d}\mu} 
    = \frac{1}{\pi} \frac{\frac{n}{2}}{\left(\frac{n}{2}\right)^2 + \mu^2},
\end{equation}

and functional $\left[ n\right]$ introduced as convolution with $K_n$:
\begin{equation}
    \left[n\right] g(\mu)\equiv K_n \star g(\mu) = \int \mathrm{d} \lambda  K_n(\mu-\lambda)g(\lambda).
\end{equation}

In terms of the string variables $\lambda^{(n)}$, the energy function can be expressed as
\begin{align}
    E=&\sum_j\frac{2\pi}{L}n_j+\sum_n D\int\mathrm{d}\lambda \sigma_n(\lambda)\left[\Theta_{n} \left(  b-\lambda\right)+\Theta_{n} \left(\lambda+  b\right)-4\pi\right].
\end{align}

Here, the first term is the charge energy
\begin{equation}
E^{(c)}(\{n_j\}) = \frac{2\pi}{L} \sum_{j=1}^{N^e} n_j,
\end{equation}
such that the charge partition function becomes
\begin{equation}
Z^{(c)} = \sum_{\{n_j\},  n_j \geq -N^e} \exp\left[-\frac{1}{T} \sum_{j=1}^{N^e} \frac{2\pi}{L} n_j \right].
\end{equation}
This describes the thermodynamics of $N$ noninteracting spinless fermions with linear kinetic energy. In the limit $D \to \infty$, it leads to the free energy
\begin{equation}
F^{(c)} = - \frac{L T}{2 \pi} \int_{-\infty}^\infty dk \ln \left(1 + e^{-\frac{k}{T}} \right) = -\frac{\pi}{12} L T^2 + \{\text{infinite constant}\}.
\end{equation}
This free energy corresponds to half the free energy of a noninteracting electron gas at zero magnetic field.

The free energy of the spin part in the presence of the magnetic field $H$ can be written as
\begin{equation}
     {F}=E+2M H - T {S},
\end{equation}
where $ {S}$ is the Yang-Yang entropy, which, upon using the Stirling approximation, can be written as
\begin{equation}
     {S}=\sum_{n=1}^\infty \int \mathrm{d} \lambda  \left[ (\sigma_n+\sigma_n^h)\ln(\sigma_n+\sigma_n^h)-\sigma_n\ln\sigma_n - \sigma_n^h\ln\sigma_n^h \right].
\end{equation}
Combining $E+M H$ as
\begin{equation}
    E+M H =\sum_{n=1}^\infty \int \mathrm{d} \lambda   g_n(\lambda) \sigma_n(\lambda),
\end{equation} 
and introducing $$g_n(\lambda)=2nH+D\left[ \Theta_{n} \left(  b-\lambda\right)+\Theta_{n} \left(\lambda+  b\right)-4\pi \right],$$ one can write the free energy as
\begin{equation}\label{Freeeng-GNK}
     {F}=\sum_{n=1}^\infty \int \mathrm{d} \lambda   \left[ g_n \sigma_n - T\sigma_n \ln\left[1+\frac{\sigma_n^h}{\sigma_n}\right] - T\sigma_n^h \ln\left[1+\frac{\sigma_n}{\sigma_n^h}\right] \right].
\end{equation}
Varying the free energy subjected to the constrain $\delta \sigma_n^h = -\sum_{m=1}^\infty A_{nm}\delta \sigma_m$  from Eq.\eqref{sigma_n^h} we get
\begin{equation}
    g_n-T\ln\left[1+\frac{\sigma_n^h}{\sigma_n}\right]+T\sum_{m=1}^\infty A_{nm}\ln\left[1+\frac{\sigma_m}{\sigma_m^h}\right]=0.
\end{equation}
or, introducing $\eta_n = \sigma_n^h/\sigma_n$, one can write
\begin{equation}\label{pre-TBA-cGNK}
\ln\left[1+\eta_n(\lambda)\right]=\frac{g_n(\lambda)}{T}+\sum_{m=1}^\infty A_{nm} \ln\left[1+\eta^{-1}_m(\lambda)\right].
\end{equation}

It is convenient to introduce a functional $G$ acting by convolution with $1/2\cosh(\pi \lambda)$
\begin{equation}
    Gf(\lambda) = \int \mathrm{d} \mu   \frac{1}{2\cosh\pi (\lambda-\mu)}f(\mu).
\end{equation}

Applying $\delta_{m,n}-G(\delta_{m-1,n}+\delta_{m+1,n})$ to the Eq.\eqref{pre-TBA-cGNK}, we obtain
\begin{equation}\label{TBA}
    \ln \eta_n(\lambda) = 
   -\frac{m}{T}\cosh(\pi \lambda)\delta_{n,1}+G\ln\left[ 1+\eta_{n+1} \right]+G\ln\left[ 1+\eta_{n-1} \right].
\end{equation}
These are the thermodynamic Bethe Ansatz equations of the model, where $\eta_n$ functions are given by a recursion relation. To complete the recursion relation, we need to impose two boundary conditions, which are $\eta_0=0$ and a relation for $n\to\infty$ of the form
\begin{equation}
		\lim_{n\to \infty}\left\{[n+1]\ln(1+\eta_n)-[n]\ln(1+\eta_{n+1})\right\}=-\frac{h}{T}.
	\end{equation}

Applying $G$ on Eq.\eqref{pre-TBA-cGNK}, we obtain
\begin{align}
    G\left[\ln[1+\eta_n(\lambda)]-\frac{g_n(\lambda)}{T} \right]=\sum_{m=1}^\infty Y_{n,m}\ln[1+\eta_n^{-1}(\lambda)]
\end{align}
where we introduced 
    \begin{equation}
        Y_{n,m}(\mu)=\sum_{l=1}^{\mathrm{min}(n,m)} K_{n+m+1-2l}(\mu).
\end{equation}
Noticing
\begin{equation}
    Y_{1,m}(\mu)=K_m(\mu),
\end{equation}
we can simplify the equation for free energy as 
\begin{equation}
    \begin{aligned}
         {F}_{\rm saddle}&=\frac{1}{2}\int \mathrm{d} \lambda  \left\{ \left( \frac{N}{2\cosh \pi (\lambda-b)}+\frac{N}{2\cosh \pi (\lambda+b)}\right)\left[ g_{1}(\lambda)-T\ln(1+\eta_{1}(\lambda)) \right]\right.\\
        &\left.+\left( \frac{1}{2\cosh\pi \lambda}+\frac{1}{2\cosh\pi (\lambda-d)}+\frac{1}{2\cosh\pi (\lambda+d)} \right)\left[g_1(\lambda)-T\ln(1+\eta_1(\lambda))\right] \right\}.
    \end{aligned}
\end{equation}
The impurity part of the free energy in the Kondo regime is
\begin{equation}
    \begin{aligned}
         {F}_{\mathrm{imp}}&= {F}_{\mathrm{imp}}^0-\frac{T}{2}\int \mathrm{d} \lambda \left(\frac{1}{2\cosh\pi (\lambda-d)}+\frac{1}{2\cosh\pi (\lambda+d)} \right)\ln(1+\eta_1(\lambda)).
    \end{aligned}
    \label{freeeneggKK}
\end{equation}

Notice that, The full free energy of the spin part receives in addition to the saddle point contribution
an extra term due to the density of states Jacobian and quadratic fluctuations around the
saddle. The exact result for the partition function is
\begin{equation}
    Z^{(s)} = e^{-\beta F_{\text{saddle}}} 
    \frac{\det(\mathbf{1}-\widehat{\mathcal Q}^{-})}{\det(\mathbf{1}-\widehat{\mathcal Q}^{+})},
\end{equation}
so that the total spin free energy is
\begin{equation}
    F^{(s)} = F_{\text{saddle}}
    -T\left[\ln\det(\mathbf{1}-\widehat{\mathcal Q}^{-})
    -\ln\det(\mathbf{1}-\widehat{\mathcal Q}^{+})\right].
\end{equation}

The operators $\widehat{\mathcal Q}^{\pm}$ act on vector functions $f=\{f_m(\mu)\}_{m\geq 1}$,
with rapidity domain $\mu>0$, as
\begin{equation}
    (\widehat{\mathcal Q}^{\pm} f)_n(\lambda)
    = \sum_{m=1}^\infty \int_0^\infty \frac{d\mu}{2\pi} 
    \frac{a_{nm}(\lambda-\mu)\ \pm\ a_{nm}(\lambda+\mu)}{1+\eta_m(\mu)} f_m(\mu),
\end{equation}
where the string--string kernel is
\begin{equation}
    a_{nm}(x)=\frac{1}{2\pi}\frac{d}{dx}\Theta_{nm}(x).
\end{equation}

Here $F_{\text{saddle}}$ denotes the saddle-point free energy, i.e. the sum of the bulk and impurity
contributions obtained above, while the determinant ratio accounts for the non-trivial
normalization in the functional integral, which gives $O(1)$ contribution to free energy independent of the impurity. The weight $1/(1+\eta_m(\mu))$ involves the
equilibrium ratios $\eta_m(\mu)=\sigma_m^h(\mu)/\sigma_m(\mu)$ determined from the TBA
equations \eqref{TBA}. By taking the ratio of the partition function of the system with and without impurity, we isolate the contribution solely from the impurity and do not explicitly compute the contributions from the two Fredholm determinants.

Deep in the Kondo phase with $d \gg 1$, define $\zeta = \pi \lambda$ and rewrite as
\begin{equation}
 {F}_{\mathrm{imp}} =  {F}_{\mathrm{imp}}^0 - \frac{T}{2 \pi} \int_{-\infty}^\infty d\zeta   K_d(\zeta) \ln\big(1 + \eta_1(\zeta/\pi)\big),
\end{equation}
where $K_d(\zeta) = \frac{1}{2 \cosh(\zeta - \pi d)} + \frac{1}{2 \cosh(\zeta + \pi d)}$.

At low temperature, $\ln(1 + \eta_1(\zeta/\pi)) \sim e^{-\frac{m}{T} \cosh \zeta}$. Near $\zeta = \pm \pi d$, setting $x = \zeta - \pi d$, and defining the Kondo temperature
\begin{equation}
T_K = \frac{m}{2} e^{\pi d},
\end{equation}
we approximate
\begin{equation}
K_d(\pi d + x) \ln\big(1 + \eta_1((\pi d + x)/\pi)\big) \approx \frac{1}{2 \cosh x} e^{-\frac{T_K}{T} e^{x}},
\end{equation}
and similarly near $-\pi d$.

Hence, the impurity-free energy correction is
\begin{equation}
\delta  {F}_{\mathrm{imp}} = - \frac{T}{2 \pi} \int_{-\infty}^\infty dx \frac{ e^{-\frac{T_K}{T} e^{x}} + e^{-\frac{T_K}{T} e^{-x}} }{2 \cosh x}.
\end{equation}

With the dimensionless variable $t = T / T_K$, the integral reduces to
\begin{equation}
I(t) = \int_{-\infty}^\infty dx \frac{ e^{-\frac{1}{t} e^{x}} + e^{-\frac{1}{t} e^{-x}} }{2 \cosh x} = 2 t \int_0^\infty dx \frac{e^{-x}}{1 + t^2 x^2}.
\end{equation}

Expanding for small $t$,
\begin{equation}
I(t) = 2 t - 4 t^3 + \mathcal{O}(t^5).
\end{equation}

Therefore,
\begin{equation}
\delta  {F}_{\mathrm{imp}} = - \frac{T}{2 \pi} I(t) = - \frac{T^2}{\pi T_K} + \frac{2 T^4}{\pi T_K^3} + \cdots,
\end{equation}
and the impurity specific heat is
\begin{equation}
C_{\mathrm{imp}}(T) = - T \frac{\partial^2 \delta  {F}_{\mathrm{imp}}}{\partial T^2} = \frac{2}{\pi} \frac{T}{T_K} + \mathcal{O}\left( \frac{T^3}{T_K^3} \right),
\end{equation} 
which shows the Fermi liquid behavior similar to the conventional Kondo problem.

\subsection{The YSR phase I}\label{sec:YSRI-tba-det}
We follow the same method as in the Kondo phase, but now, we need to construct the towers of excitations differently for the states built out of the base state $\ket{U}$ and the base state $\ket{K}$. For the base state $\ket{U}$, we plug in the string solutions and write down the Bethe Ansatz equations in logarithmic form as

\begin{equation}
  \Theta_{n}(\lambda^{(n)}_\gamma)+\sum_\upsilon  \Theta_{n+2\upsilon  \delta}(\lambda^{(n)}_\gamma)+N \Theta_{n}(\lambda^{(n)}_\gamma+\upsilon  b)=\sum_{m,\beta}\Theta_{n,m}(\lambda^{(n)}_\gamma+\upsilon \lambda^{(m)}_\beta)-2\pi I_\gamma^{(n)},
\end{equation}

As shown explicitly in the Kondo phase, we carry out the usual TBA procedure and arrive at the impurity free energy of the form
    \begin{align}
        {F}_{\rm imp}^{(\mathcal{T}_1)}= {F}_{0}-\frac{T}{2}\sum_n \int \mathrm d \lambda K_{n+2\upsilon\gamma}(\lambda) \ln \left[1+\eta^{-1}_n(\lambda)\right]
        \label{T2-YSR1}
    \end{align}

Likewise, for the towers of excited states above the state $\ket{B}$, plugging in the string solutions, we obtain Bethe Ansatz equations of the form
\begin{equation}
  \Theta_{n}(\lambda^{(n)}_\gamma)-\sum_\upsilon  \Theta_{n+2\upsilon  (\delta-1)}(\lambda^{(n)}_\gamma)+N \Theta_{n}(\lambda^{(n)}_\gamma+\upsilon  b)=\sum_{m,\beta}\Theta_{n,m}(\lambda^{(n)}_\gamma+\upsilon \lambda^{(m)}_\beta)-2\pi J_\gamma^{(n)},
\end{equation}
which results in 
\begin{align}
    {F}_{\rm imp}^{(\mathcal{T}_2)}= {F}_{0}+\frac{T}{2}\sum_{n,\upsilon} \int \mathrm{d}  \lambda K_{n+\upsilon 2(\delta-1)}(\lambda) \ln \left[1+\eta^{-1}_n(\lambda)\right]
    \label{F2inveqn}
\end{align}

Notice that Eq.\eqref{pre-TBA-cGNK} admits an analytic continuation to the strip in the complex plane given by $\lambda \pm i \zeta$ when $|\zeta|<\frac12$
\begin{equation}\label{pre-TBA-cGNK-C}
    \sum_{\upsilon=\pm} G\left[ \ln \left(1+\eta_m(\lambda+\upsilon i\zeta)\right) - \frac{g_m(\lambda+\upsilon i\zeta)}{T} \right]
    = \sum_{\upsilon=\pm} \sum_n Y_{mn} * \ln \left(1+\eta_n^{-1}(\lambda+\upsilon i\zeta)\right) .
\end{equation}

Thus, using Eq.\eqref{pre-TBA-cGNK} and simplifying, one arrives at
\begin{align}
    F_{\rm imp}^{(\mathcal{T}_1)}&=F_{\rm imp}^{(\mathcal{T}_2)}+|E_\delta|-\int \rm{d} \lambda \sum_{\upsilon=\pm}\frac{\frac{T}{4}\ln \left[1+\eta_2(\lambda)\right]}{\cosh(\pi(\lambda+i\upsilon (\delta-\frac12)))}\\
    F_{\rm imp}^{(\mathcal{T}_2)}&=-\frac{T}{4}\int \rm{d} \lambda \sum_{\upsilon=\pm}\frac{\ln \left[1+\eta_1(\lambda)\right]}{\cosh(\pi(\lambda+i\upsilon \delta))}
\end{align}
where $E_\delta=-m\sin(\pi\delta)$ is the spin part of the energy of the localized YSR mode. To show the explicit computation, in order to simplify the $T$ dependent terms in the integral in Eq.\eqref{F2inveqn}, we perform the following steps
\begin{align}
    \mathcal{I}&=\frac{T}{2}\sum_{n,\upsilon} \int \mathrm{d}  \lambda K_{n+\upsilon 2(\delta-1)}(\lambda) \ln \left[1+\eta^{-1}_n(\lambda)\right]\nonumber\\
    &=\frac{T}{2}\sum_{n,\upsilon} \int \mathrm{d}  \lambda K_{n}(\lambda+i\upsilon (\delta-1)) \ln \left[1+\eta^{-1}_n(\lambda)\right]\nonumber\\
    &=\frac{T}{2}\sum_{n,\upsilon=\pm} \int \mathrm{d}\lambda 
K_{n}(\lambda) 
\ln \left[1+\eta_n^{-1} \bigl(\lambda - i\upsilon(\delta-1)\bigr)\right]
\end{align}

Using Eq.\eqref{pre-TBA-cGNK-C}, we can thus write
\begin{align}
    F_{\rm imp}^{(\mathcal{T}_2)}-F_0^{(\mathcal{T}_2)}&=\frac{T}{4} \sum_{\upsilon=\pm} 
\frac{\ln \left( 1 + \eta_1(\lambda) \right)}
{\cosh \big[ \pi\big(\lambda + i\upsilon (\delta-1)\big) \big]}\nonumber\\
&=\frac{T}{4} \int \mathrm{d}\lambda \sum_{\upsilon=\pm} 
\frac{\ln \left[1+\eta_1(\lambda)\right]}
{\cosh \big(\pi(\lambda+i\upsilon \delta)\big)}.
\end{align}
Where $F_0^{(\mathcal{T}_2)}$ is the energy of the state $\ket{U}$. A similar procedure can be used to simplify Eq.\eqref{T2-YSR1} by always shifting the variable such that the complex part of the argument is less than $\frac{1}{2}$ and hence Eq.\eqref{pre-TBA-cGNK-C} can be used. Upon such calculations, one arrives at
\begin{align}
    F_{\rm imp}^{(\mathcal{T}_1)}-F_0^{(\mathcal{T}_1)}&=F_{\rm imp}^{(\mathcal{T}_2)}+|E_\delta|-\int \rm{d} \lambda \sum_{\upsilon=\pm}\frac{\frac{T}{4}\ln \left[1+\eta_2(\lambda)\right]}{\cosh(\pi(\lambda+i\upsilon (\delta-\frac12)))}.
\end{align}

Upon noting that constant ground-state energies can be absorbed into the partition function, 
we set $F_{0}^{(\mathcal{T}_2)} = 0$. 
With this choice, the energy difference between the $\ket{B}$ and $\ket{K}$ states, 
$|E_\delta|$, is identified with $F_{0}^{(\mathcal{T}_2)}$, thereby completing the derivation of the above equations.

Notice that in the Kondo phase only $\eta_1$ contributes to the free energy of the impurity, but in the YSR I phase both $\eta_1$ and $\eta_2$ contribute to the total free energy of the impurity. 

\begin{figure}[H]
    \centering
\includegraphics[width=0.75\linewidth]{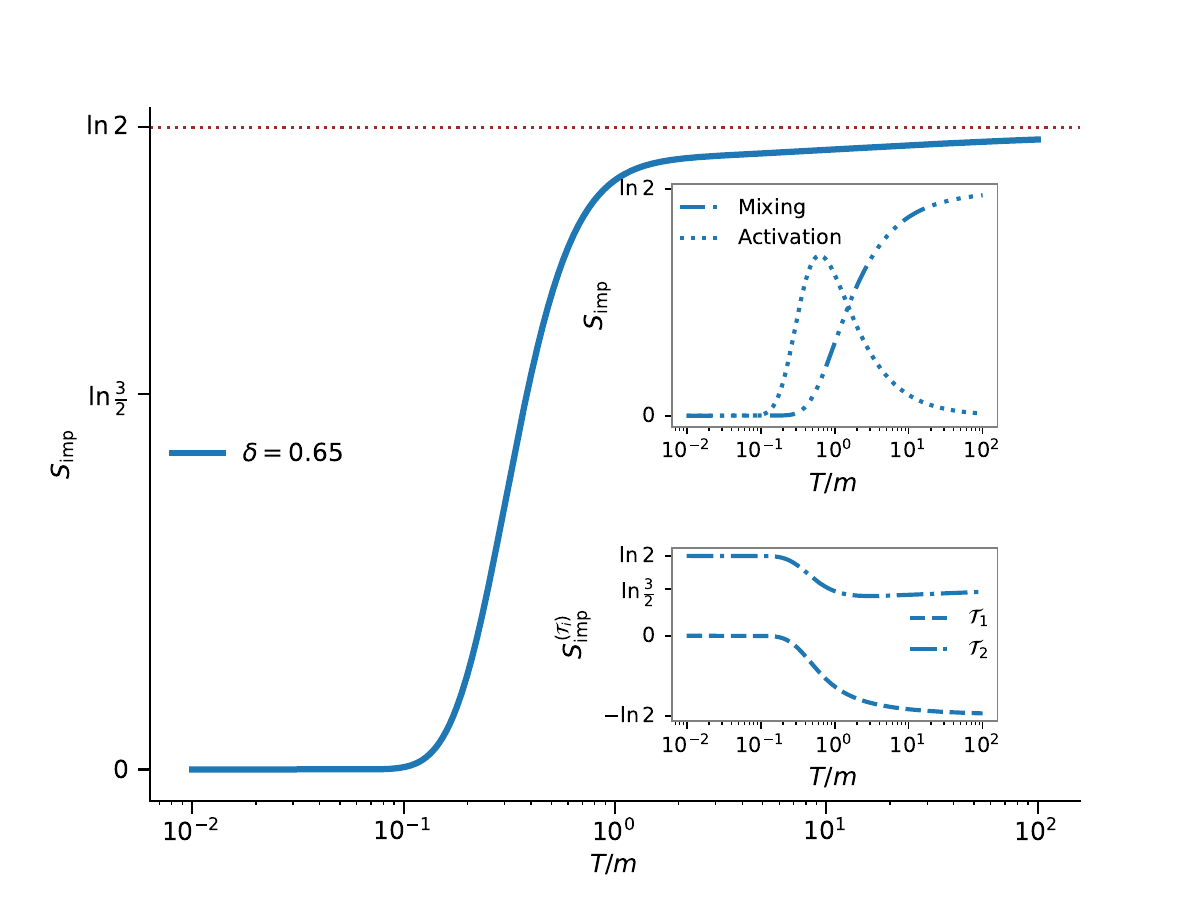}
    \caption{The impurity entropy contribution for $\delta=0.65$ is shown in the main panel. The top-right inset displays the \emph{mixing term}, representing the weighted ensemble average of the two towers. This term increases monotonically from zero at zero temperature to $\ln 2$ at high temperature, reflecting the gradual thermal population of degrees of freedom. The inset also shows the \emph{activation term}, which becomes significant only near the temperature scale $T \sim E_\delta$, corresponding to the thermal excitation of the midgap state. Notice that due to the slow growth of the mixing term, the bump due to the activation term is not enough to bring the entropy higher than $\ln2$ for smaller values of $\delta$ but as shown in main text, as $\delta$ becomes closer to 1, then the impurity entropy can onvershoot the free moment value of $\ln 2$ at intermediate temperatures. The bottom-right inset presents the individual entropy contributions of each tower. The first tower’s entropy flows from $0$ at zero temperature—where the impurity is fully screened—to $-\ln 2$ at infinite temperature, indicating the removal of a bulk degree of freedom due to singlet formation with the impurity. Conversely, the second tower starts at $\ln 2$ at zero temperature and approaches $3/2$ at high temperature, signifying the formation of a triplet state between impurity and bulk degrees of freedom.
}
    \label{fig:delsmallYSRI}
\end{figure}

As illustrated in Fig.~\ref{fig:delsmallYSRI}, for $\delta=0.65$, the impurity is fully screened at low temperature, causing the impurity entropy to vanish, while at infinite temperature it becomes asymptotically free, recovering the free moment value $\ln 2$. The free energy receives contributions from two distinct towers: their weighted ensemble average—the mixing term—yields a monotonic increase of the impurity entropy from zero at zero temperature to $\ln 2$ at high temperature. Meanwhile, the activation term, associated with the thermal occupation of the midgap Yu-Shiba-Rusinov (YSR) state at energy $|E_\delta|$, produces a sharp peak around $T \sim |E_\delta|$. For $\delta$ closer to 0.5, the mixing term grows slowly, and the activation bump is insufficient to push the total entropy above $\ln 2$ at intermediate temperatures. As $\delta$ approaches 1, the more rapid rise of the mixing term causes the impurity entropy to exceed $\ln 2$ transiently. The lower-right inset decomposes the entropy into tower contributions: tower 1 corresponds to a screened impurity at low temperature with zero entropy, and at high temperature removes one bulk degree of freedom due to singlet formation, contributing $-\ln 2$; tower 2 features an unscreened impurity at low temperature contributing $\ln 2$, evolving to a triplet state with bulk degrees of freedom at high temperature, thereby contributing $\ln \frac{3}{2}$ to the entropy.

\subsection{YSR phase II}\label{sec:YSRIItherm}

As shown in Sec.\ref{sec:YSRII-gs}, there are three distinct towers in this phase. Following the same procedure as above, we arrive at the contribution from the three towers $\mathcal{T}_1$ built on top of the state $\ket{U}$, $\mathcal{T}_2$ built on top of state $\ket{B}$ and finally $\mathcal{T}_3$ built on top of state $\ket{\tilde U}$ as

\begin{align}
F_{\mathrm{imp}}^{(\mathcal{T}_1)} &= -\frac{T}{2} \sum_{n} \sum_{\upsilon = \pm} \int \mathrm{d}\lambda   K_{n + 2 \upsilon \delta}(\lambda) \ln \left[ 1 + \eta_n^{-1}(\lambda) \right], \\
F_{\mathrm{imp}}^{(\mathcal{T}_2)} &= \frac{T}{2} \sum_{n} \sum_{\upsilon = \pm} \int \mathrm{d}\lambda   K_{n + 2 \upsilon (\delta - 1)}(\lambda) \ln \left[ 1 + \eta_n^{-1}(\lambda) \right], \\
F_{\mathrm{imp}}^{(\mathcal{T}_3)} &= \frac{T}{2} \sum_{n} \sum_{\upsilon = \pm} \int \mathrm{d}\lambda   \Big[ K_{n + 2 \upsilon (\delta - 2)}(\lambda) + 2 K_{n + 2 \upsilon (\delta - 1)}(\lambda) \Big] \ln \left[ 1 + \eta_n^{-1}(\lambda) \right].
\end{align}

Once again using Eq.\eqref{pre-TBA-cGNK}, we can rewrite these equations in the form
\begin{align}
F_{\rm imp}^{(\mathcal{T}_1)} &= -\frac{T}{4} \int_{-\infty}^\infty \mathrm{d}\lambda \sum_{\upsilon=\pm} \frac{\ln\bigl(1+\eta_3(\lambda)\bigr)}{\cosh\left(\pi(\lambda + i \upsilon (\delta - 1))\right)} + \frac{T}{4} \int_{-\infty}^\infty \mathrm{d}\lambda \sum_{\upsilon=\pm} \frac{\ln\bigl(1+\eta_2(\lambda)\bigr)}{\cosh\left(\pi(\lambda + i \upsilon (\tfrac{3}{2} - \delta))\right)}, \\
F_{\rm imp}^{(\mathcal{T}_2)} &= E_\delta + \frac{T}{4} \int_{-\infty}^\infty \mathrm{d}\lambda \sum_{\upsilon=\pm} \frac{\ln\bigl(1+\eta_1(\lambda)\bigr)}{\cosh\left(\pi(\lambda + i \upsilon (\delta - 1))\right)}, \\
F_{\rm imp}^{(\mathcal{T}_3)} &= \frac{T}{4} \int_{-\infty}^\infty \mathrm{d}\lambda \sum_{\upsilon=\pm} \frac{\ln\bigl(1+\eta_2(\lambda)\bigr)}{\cosh\left(\pi(\lambda + i \upsilon (\tfrac{3}{2} - \delta))\right)} + \frac{T}{4} \int_{-\infty}^\infty \mathrm{d}\lambda \sum_{\upsilon=\pm} \frac{\ln\bigl(1+\eta_1(\lambda)\bigr)}{\cosh\left(\pi(\lambda + i \upsilon (\delta - 1))\right)}.
\end{align}

\begin{figure}
    \centering    \includegraphics[width=0.75\linewidth]{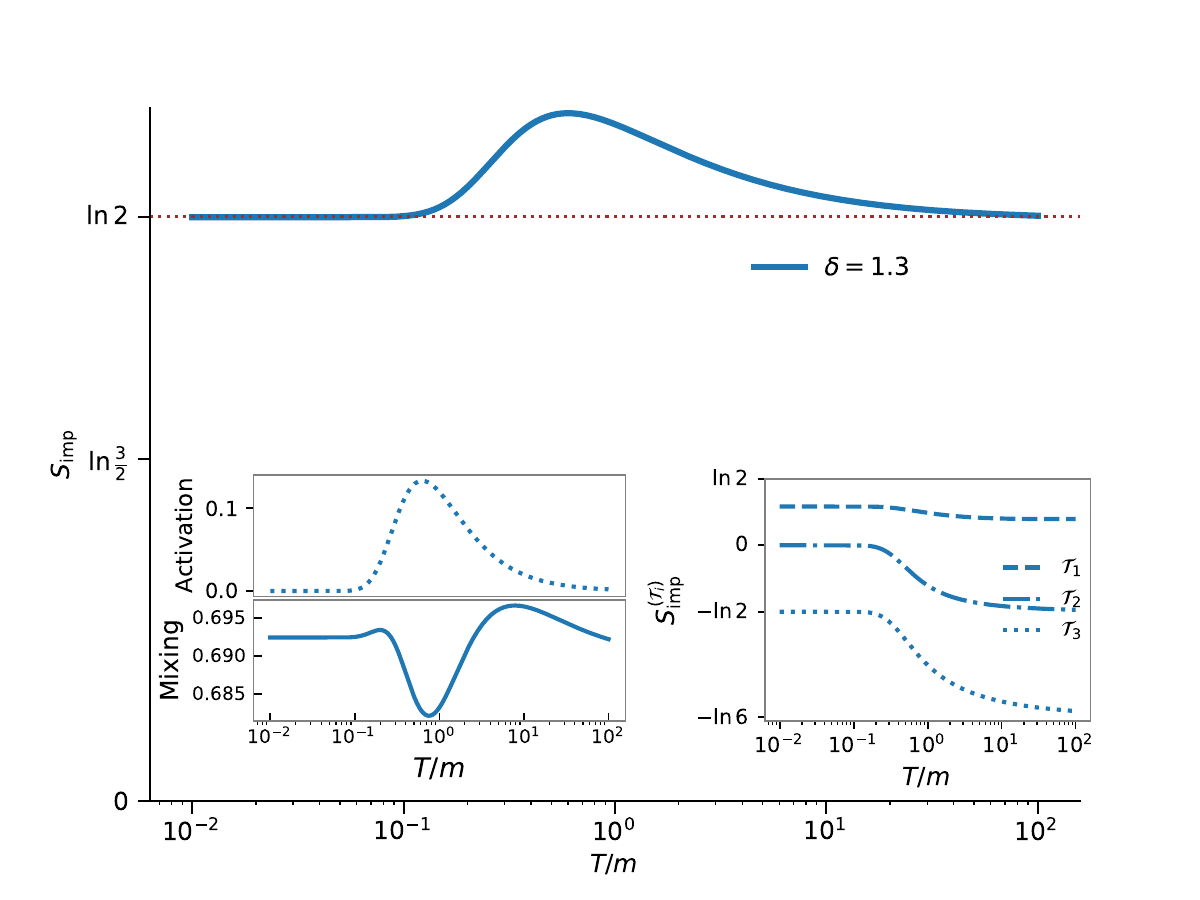}
    \caption{Representative impurity entropy $S_{\rm imp}(T)$ for $\delta = 1.3$.
        The main panel shows the total impurity entropy, which approaches $\ln 2$ in both the low-temperature ($T \to 0$) and high-temperature ($T \to \infty$) limits, but exhibits a pronounced bump at intermediate temperatures. 
        The right inset displays the entropy contributions from the three individual towers $\mathcal{T}_1$, $\mathcal{T}_2$, and $\mathcal{T}_3$.
        The left inset depicts the mixing term, defined as the weighted ensemble average of the three tower contributions, alongside the activation term, which is significant only near $T \sim |E_\delta|$. 
        The activation term reflects the thermal activation of the midgap state contributing to the entropy enhancement in the intermediate temperature regime.
    }
    \label{fig:del1p3}
\end{figure}

As illustrated in Fig.~\ref{fig:del1p3} for the representative case of $\delta = 1.3$, the total impurity contribution to the free energy in the unscreened regime approaches $\ln 2$ in both the low-temperature $(T \to 0)$ and high-temperature $(T \to \infty)$ limits. However, in the intermediate temperature range, the impurity entropy exceeds $\ln 2$. 

The right inset of the figure shows the contributions from each tower, which  at low temperature take the following values 
\begin{equation}
S_{\rm imp}^{(\mathcal{T}_1)}(T=0) = \ln \frac{3}{2}, \quad
S_{\rm imp}^{(\mathcal{T}_2)}(T=0) = 0, \quad
S_{\rm imp}^{(\mathcal{T}_3)}(T=0) = -\ln 2.
\end{equation}

Similarly, in the high-temperature limit, the entropies are given by
\begin{equation}
S_{\rm imp}^{(\mathcal{T}_1)}(T \to \infty) = \ln \frac{4}{3}, \quad
S_{\rm imp}^{(\mathcal{T}_2)}(T \to \infty) = -\ln 2, \quad
S_{\rm imp}^{(\mathcal{T}_3)}(T \to \infty) = -\ln 6.
\end{equation}
  
At zero temperature, only $\mathcal{T}_1$ and $\mathcal{T}_3$ contribute as $\mathcal{T}_3$ has an energy difference of $E_\delta>0$. Thus, the zero temperature entropy becomes $S_{\rm imp}(T=0)=\ln(e^{S_{\rm imp}^{(\mathcal{T}_1)}(T=0)}+e^{S_{\rm imp}^{(\mathcal{T}_3)}(T=0)})=\ln 2$. At infinite temperature, all the towers equally contribute, leading to the infinite temperature total entropy of the impurity to be $S_{\rm imp}(T\to\infty)=\ln\left( e^{S_{\rm imp}^{(\mathcal{T}_1)}(T \to \infty)}+e^{S_{\rm imp}^{(\mathcal{T}_2)}(T \to \infty)}+e^{S_{\rm imp}^{(\mathcal{T}_3)}(T \to \infty)}\right)$. The total entropy due to the three towers for any temperature $T$ can be written as
\begin{equation}
    S_{\rm imp}(T) 
= \ln \left(
    e^{S_{\rm imp}^{(\mathcal{T}_1)}} 
    + e^{S_{\rm imp}^{(\mathcal{T}_2)} - \beta E_\delta} 
    + e^{S_{\rm imp}^{(\mathcal{T}_3)}}
\right) 
+ \frac{E_\delta}{T}  
\frac{
    e^{S_{\rm imp}^{(\mathcal{T}_2)} - \beta E_\delta}
}{
    e^{S_{\rm imp}^{(\mathcal{T}_1)}} 
    + e^{S_{\rm imp}^{(\mathcal{T}_2)} - \beta E_\delta} 
    + e^{S_{\rm imp}^{(\mathcal{T}_3)}}
},
\end{equation}
where the first term is the weighted ensemble average of the three towers that we call the mixing term, which remains close to $\ln 2$ with some dips and bumps as shown in the left inset, whereas the second term is the activation term which is only nonvanishing when $T\sim E_\delta$ is the contribution from the thermal activation of the midgap state. Together, these two terms give the total contribution to the impurity entropy.

\subsection{Uscreened phase}
Once again, we write the Bethe equations for $n-$strings on top of the base states in each of the towers and follow the same procedure as above to arrive at the free energy equations:
		\begin{align}
			F_{\rm imp}^{(\mathcal{T}_1)} &= - \frac{T}{4} \sum_{\upsilon = \pm} \int \mathrm{d}\lambda \left[\frac{\ln\left(1 + \eta_{\lceil 2\delta \rceil}(\lambda)\right)}{ 
				\cosh\left(\pi\left(\lambda + \frac{i\upsilon}{2}(2\delta - \lfloor 2\delta \rfloor)\right)\right)} 
			-\frac{\ln\left(1 + \eta_{\lfloor 2\delta \rfloor}(\lambda)\right)}{\cosh\left(\pi\left(\lambda + \frac{i\upsilon}{2}(\lceil 2\delta \rceil - 2\delta)\right)\right)} 
			\right], \nonumber\\
			F_{\rm imp}^{(\mathcal{T}_2)} &= \frac{T}{4} \sum_{\upsilon = \pm} \int \mathrm{d}\lambda \left[ 
			\frac{\ln\left(1 + \eta_{\lceil 2\delta \rceil - 2}(\lambda)\right)}{\cosh\left(\pi\left(\lambda + \frac{i\upsilon}{2}(2\delta - \lfloor 2\delta \rfloor)\right)\right) }
			- \frac{\ln\left(1 + \eta_{\lfloor 2\delta \rfloor - 2}(\lambda)\right)}{\cosh\left(\pi\left(\lambda + \frac{i\upsilon}{2}(\lceil 2\delta \rceil - 2\delta)\right)\right) }
			\right], \nonumber\\
			F_{\rm imp}^{(\mathcal{T}_3)} &= \frac{T}{4} \sum_{\upsilon = \pm} \int \mathrm{d}\lambda \left[ \frac{\ln\left(1 + \eta_{\lfloor 2\delta \rfloor}(\lambda)\right)}{\cosh\left(\pi\left(\lambda + \frac{i\upsilon}{2}(\lceil 2\delta \rceil - 2\delta)\right)\right)}
			+ \frac{\ln\left(1 + \eta_{\lceil 2\delta \rceil - 2}(\lambda)\right)}{\cosh\left(\pi\left(\lambda + \frac{i\upsilon}{2}(2\delta - \lfloor 2\delta \rfloor)\right)\right)}
			\right].\nonumber
		\end{align}

\begin{figure}[H]
    \centering
    \includegraphics[width=0.75\linewidth]{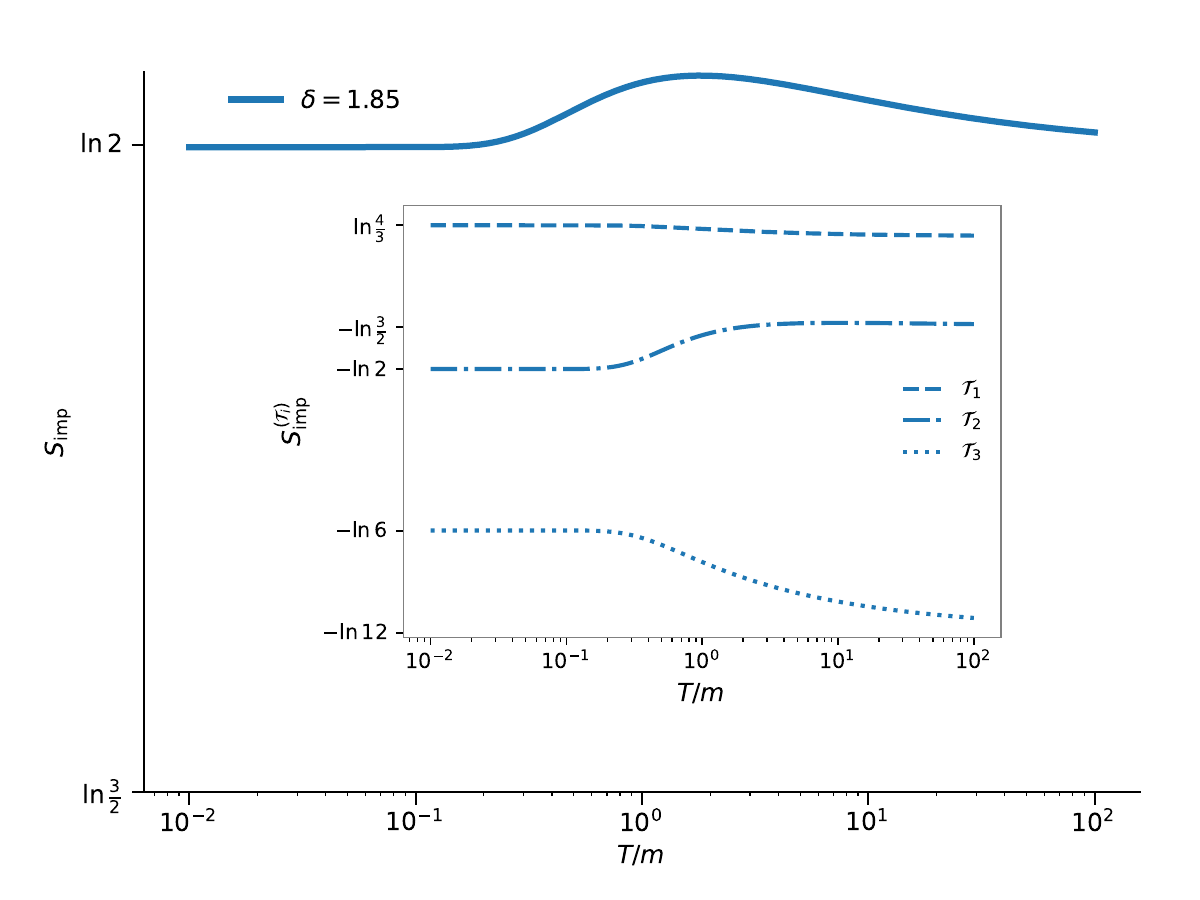}
    \caption{Impurity entropy $ S_{\mathrm{imp}} $ as a function of the scaled temperature $ T/m $ for $ \delta = 1.85 $. 
        The main panel shows the total impurity entropy exhibiting non-monotonic temperature dependence with a peak slightly above $\ln 2$.
        The inset displays the individual entropy contributions from the three towers $ S_{\mathrm{imp}}^{(\mathcal{T}_1)} $ (dashed), 
        $ S_{\mathrm{imp}}^{(\mathcal{T}_2)} $ (dash-dotted), and $ S_{\mathrm{imp}}^{(\mathcal{T}_3)} $ (dotted) on a logarithmic temperature scale, illustrating their distinct asymptotic behaviors at low and high temperatures.
    }
    \label{fig:del1p85pm}
\end{figure}

For the representative case of $\delta=1.85$, as shown in Fig.\ref{fig:del1p85pm}, the impurity is unscreened in both the $T\to 0$ and $T\to\infty$ limits. And there is some intermediate bump in the entropy such that the impurity entropy exceeds $\ln 2$. Notice that there are no mid-gap states in this phase, but despite this, the intricate interplay between the three towers leads to this overshoot in the value of the impurity entropy more than its free moment value of $\ln 2$. As shown in the inset, the individual contributions from the three towers in the two extreme limits are
\[
S_{\rm imp}^{(\mathcal{T}_1)}(0) = \ln \frac{4}{3}, \quad
S_{\rm imp}^{(\mathcal{T}_2)}(0) = \ln \frac{1}{2}, \quad
S_{\rm imp}^{(\mathcal{T}_3)}(0) = \ln \frac{1}{6},
\]
and
\[
S_{\rm imp}^{(\mathcal{T}_1)}(\infty) = \ln \frac{5}{4}, \quad
S_{\rm imp}^{(\mathcal{T}_2)}(\infty) = \ln \frac{2}{3}, \quad
S_{\rm imp}^{(\mathcal{T}_3)}(\infty) = \ln \frac{1}{12}.
\], 
and since none of the towers are lifted in energies, all three of them contribute equally in all temperature scales, such that the total entropy is given by the weighted ensemble average of the three towers i.e.
\begin{equation}
      S_{\rm imp}(T) 
= \ln \left(
    e^{S_{\rm imp}^{(\mathcal{T}_1)}} 
    + e^{S_{\rm imp}^{(\mathcal{T}_2)}} 
    + e^{S_{\rm imp}^{(\mathcal{T}_3)}}
\right).
\end{equation}

\section{Numerical Computation of the $\eta_n(\lambda)$ Functions from the TBA Equations}

We numerically solve the truncated TBA integral equations for the auxiliary functions $\eta_n(\lambda)$, with $n = 1, \ldots, n_{\max} + 1$, defined on a finite rapidity domain $\lambda \in [-L, L]$. The rapidity interval is discretized uniformly into $N_\lambda$ points, $\{\lambda_i\}_{i=1}^{N_\lambda}$, with spacing $\Delta \lambda = \frac{2L}{N_\lambda - 1}$. In our computations, we use $L = 40.0$, $N_\lambda = 1024$, and truncate the string index at $n_{\max} = 30$. These parameters control the resolution and accuracy of the numerical solution.

The functions $\eta_n(\lambda)$ satisfy the coupled nonlinear integral equations
\begin{equation}
\ln \eta_n(\lambda) = \delta_{n, n_{\mathrm{imp}}} d(\lambda) + (G \ln(1 + \eta_{n-1}))(\lambda) + (G \ln(1 + \eta_{n+1}))(\lambda),
\label{eq:tba_iteration}
\end{equation}
where the driving term $d(\lambda)$ acts only at the impurity string index $n_{\mathrm{imp}}$, and $\eta_0(\lambda) \equiv 0$. The integral convolution operator $G$ is defined by
\begin{equation}
(G f)(\lambda_i) = \int_{-L}^L \frac{f(\lambda')}{2 \cosh\left(\pi (\lambda_i - \lambda')\right)}   d\lambda',
\end{equation}
discretized with trapezoidal quadrature supplemented by an analytic tail correction to account for contributions from $|\lambda| > L$. This tail correction relies on an analytic expression for the integral kernel's contribution outside the finite domain, improving numerical accuracy without requiring larger cutoffs.

Truncation of the infinite hierarchy of strings at $n = n_{\max}$ necessitates an asymptotic closure condition for $\eta_{n_{\max} + 1}(\lambda)$, imposed via integral operators $[n]$ defined by
\begin{equation}
([n] f)(\lambda_i) = \frac{n \pi}{2} \int_{-L}^L \frac{f(\lambda')}{\pi \left( \left(\frac{n \pi}{2}\right)^2 + (\lambda_i - \lambda')^2 \right)}   d\lambda',
\end{equation}
also discretized with trapezoidal quadrature. The closure relation
\begin{equation}
[n_{\max} + 1] \ln(1 + \eta_{n_{\max}}) - [n_{\max}] \ln(1 + \eta_{n_{\max} - 1}) = -\frac{H}{T} + [n_{\max} + 1] \ln(1 + \eta_{n_{\max} + 1}),
\end{equation}
is solved explicitly for $\ln(1 + \eta_{n_{\max} + 1}(\lambda))$, which is then exponentiated and shifted to determine $\eta_{n_{\max} + 1}(\lambda)$.

The iterative procedure starts with initial guesses $\eta_0(\lambda) = 0$ and $\eta_n(\lambda) = (n+1)^2 - 1$ for $n \geq 1$, reflecting asymptotic string behavior. At each iteration, $\eta_n(\lambda)$ for $1 \leq n \leq n_{\max}$ are updated by evaluating the convolution terms using the previous iterate values, adding the driving term at $n = n_{\mathrm{imp}}$, and enforcing the asymptotic closure at $n_{\max} + 1$. The iteration continues until the maximum absolute change across all $\eta_n(\lambda)$ falls below a tolerance $\epsilon = 10^{-10}$.

The choice of the rapidity cutoff $L = 40.0$ and discretization $N_\lambda = 1024$ is motivated by the rapid decay of the convolution kernel and the auxiliary functions $\eta_n(\lambda)$, as well as by practical numerical convergence tests. Although no formal error bounds have been established, the parameters $L$ and $N_\lambda$ are chosen based on empirical convergence tests, increasing them until key computed quantities—such as the impurity free energy and entropy—vary negligibly within the target numerical precision (approximately $10^{-10}$) and agree with known analytical results in accessible limiting cases.
The analytic tail correction further improves the approximation of integrals truncated at $\pm L$ by accounting for the kernel contributions beyond the finite numerical domain, reducing truncation artifacts and enhancing numerical stability. This approach provides a robust and reproducible method for solving the TBA equations with high practical accuracy.
 The plots of the $\eta_n(\lambda)$ for $n=1\cdots 5$ are shown in the main text for a few temperature values.

\section{Solving the model with fixed particle number}
In the main text and in Sec.~\ref{sec:towerscGNK}, we solved the model in the YSR I and YSR II by allowing the particle number to change and found that at $\delta=1$, the model undergoes parity changing phase transition where the impurity is screened for $\delta<1$ and unscreened for $\delta>1$. In this section, we briefly solve the model by fixing the total number of electrons $N$ in the bulk to be odd. Notice that the $\ket{B}$ state constructed in Sec.\ref{sec:towerscGNK} is still valid as the total number of roots $M_{\ket{B}}=\frac{N+1}{2}$ is an integer. However, the state $\ket{U}$ is not a valid state as the total number of roots $M_{\ket{U}}=\frac{N}{2}$ is not an integer as $N$ is odd. Thus, we need to add a hole to construct the valid state. Upon doing so, the energy difference between the state $\ket{U}$ and $\ket{B}$ becomes
\begin{equation}
    E_{\ket{U}}-E_{\ket{B}}=m+m\cos(\pi \delta)>1.
\end{equation}
Since the addition of a massive hole shifts the energy of the $\ket{U}$ state by $m$, throughout the YSR phase (I and II), the state $\ket{B}$ is lower in energy and hence the impurity is screened in the ground state throughout this phase. 

When $\frac{1}{2}<\delta<1$, there are two towers, and the contribution to the impurity entropy from individual towers becomes

\begin{align}
		F_{\rm imp}^{(\mathcal{T}_1)}&=-\frac{T}{4}\int \rm{d} \lambda \sum_{\upsilon=\pm}\frac{\ln \left[1+\eta_1(\lambda)\right]}{\cosh(\pi(\lambda+i\upsilon \delta))} ;\nonumber\\
		F_{\rm imp}^{(\mathcal{T}_2)}&=F_{\rm imp}^{(\mathcal{T}_1)}+|E_\delta|+m-\int \rm{d} \lambda \sum_{\upsilon=\pm}\frac{\frac{T}{4}\ln \left[1+\eta_2(\lambda)\right]}{\cosh(\pi(\lambda+i\upsilon (\delta-\frac12)))},\nonumber
	\end{align}
where the only change is the shift in the energy of the second tower by additional values of the mass of a single hole $m$. 

However, this induces a significant qualitative change in the impurity entropy behavior, particularly for values of $\delta$ near the Kondo-YSR I phase boundary at $\delta = 0.5$, where the impurity entropy can exhibit a dip below zero at intermediate temperatures whereas for larger values of $\delta$ closer to $\delta=1$, the impurity entropy exceeds the free moment value of $\ln2$ in the intermediate temperatures. 
\begin{figure}[H]
		\centering
		\includegraphics[width=0.48\linewidth]{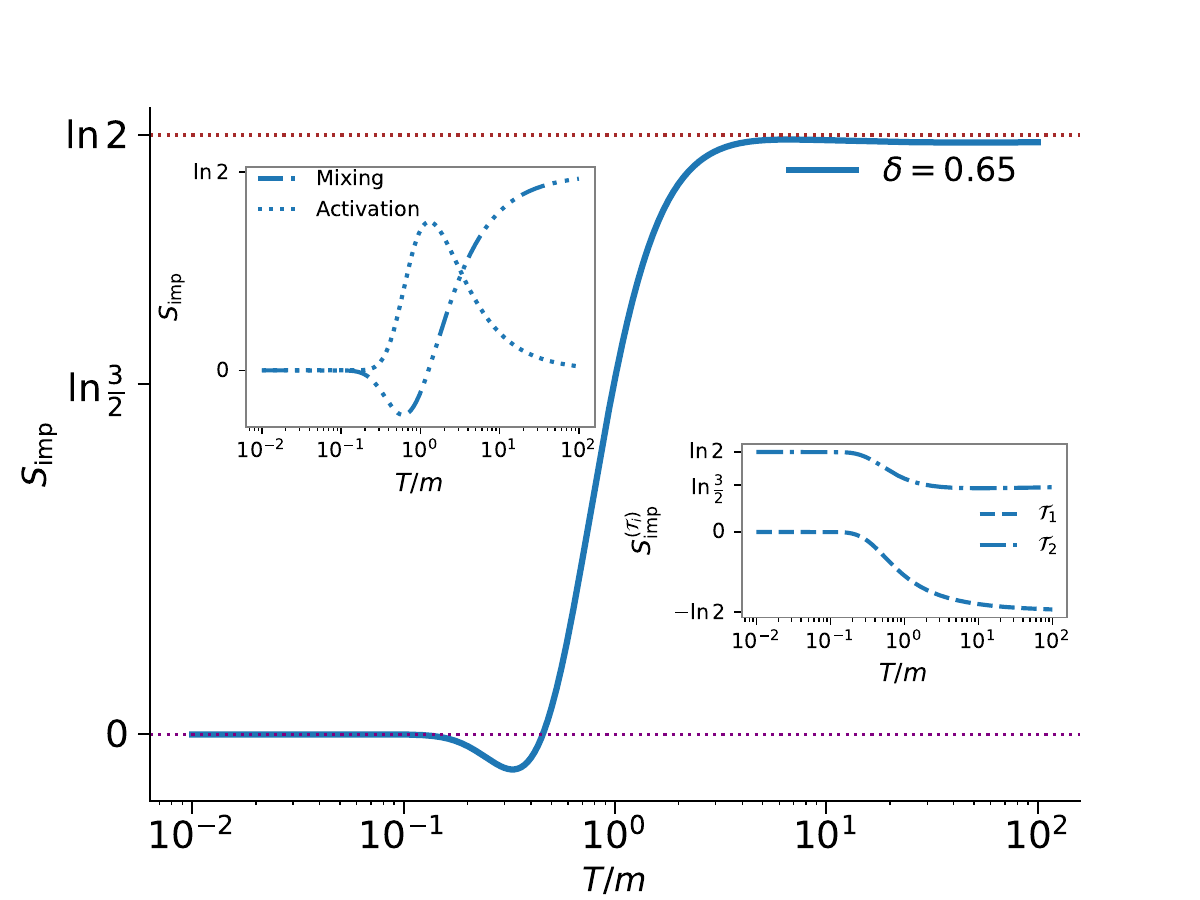}
		\includegraphics[width=0.48\linewidth]{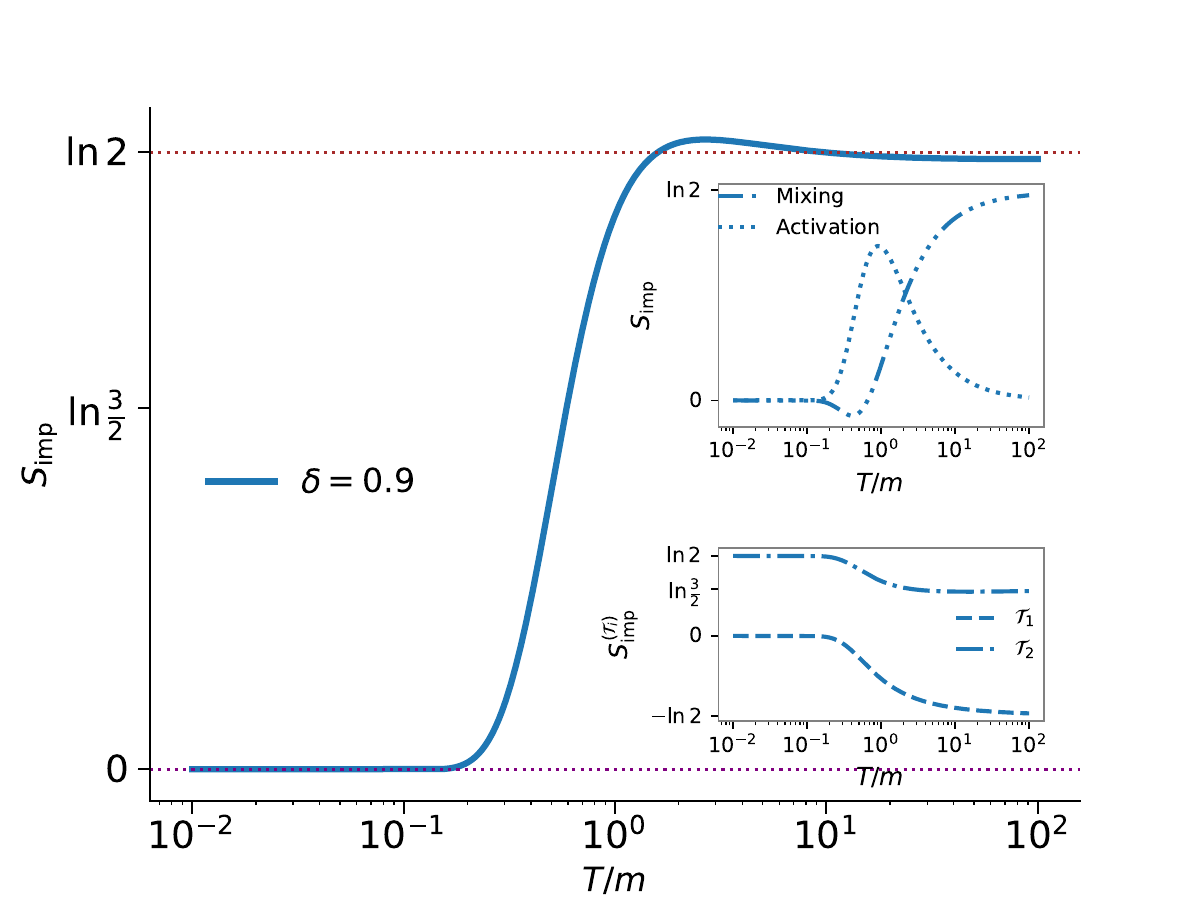}
		\caption{
			Temperature dependence of the impurity entropy $S_{\rm imp}(T)$ for two representative values of $\delta$, illustrating distinct intermediate behaviors.  \textbf{Left panel} ($\delta=0.65$): The impurity is screened at zero temperature, with the entropy vanishing as $T \to 0$, and becomes unscreened in the high-temperature limit, where the entropy saturates at $\ln 2$. A notable dip appears at intermediate temperatures below $T \sim m$. The right inset displays the entropy contributions from the two relevant towers $\mathcal{T}_1$ and $\mathcal{T}_2$, while the left inset shows the ensemble average mixing term together with the activation term associated with thermal population of the midgap state. The activation term is significant near $T \sim |E_\delta|+m$, driving entropy changes in the intermediate regime however, unlike in Sec.\ref{sec:YSRI-tba-det}, it is not enough to overcome the negative values from $\mathcal{T}_1$ where one bulk degree of freedom forms sa inglet with the impurity thereby resulting in negative entropy at intermediate temepratures. \textbf{Right panel} ($\delta=0.9$): Here, also the impurity remains screened at zero temperature, with vanishing entropy, and is unscreened at high temperature with entropy $\ln 2$, but exhibits an overshoot beyond $\ln 2$ at intermediate temperatures. The right inset shows the entropy contributions from the two towers $\mathcal{T}_1$ and $\mathcal{T}_2$, and the left inset depicts the mixing and activation terms, emphasizing the role of thermal population effects around $T \sim |E_\delta|+m$. Overall, these panels demonstrate how impurity screening and thermal activation interplay to shape the entropy profile, with the dips and overshoots in intermediate temperatures depending on $\delta$.
		}
		\label{fig:delta_0p65_0p9_entropy}
	\end{figure}

As shown in Fig.~\ref{fig:delta_0p65_0p9_entropy}, the impurity entropy $S_{\rm imp}(T)$ exhibits distinctive temperature-dependent behavior governed by the parameter $\delta$. For $\delta$ values approaching the phase boundary of the Kondo-YSR I phase ($\delta_c = 0.5$), the entropy vanishes as $T \to 0$ and saturates to $\ln 2$ at high temperatures, displaying a characteristic dip at intermediate temperatures. 
This behavior can be understood by examining the contributions from the individual towers. Specifically, if only the tower $\mathcal{T}_1$ were present, the entropy difference would approach $-\ln 2$ at infinite temperature. This reflects the formation of a singlet state between the impurity and one bulk degree of freedom (DOF), effectively reducing one degree of freedom in the bulk. In contrast, the tower $\mathcal{T}_2$ involves a triplet formation between the impurity and bulk DOFs, such that the difference in entropy at infinite temperature reaches $\ln \frac{3}{2}$. At intermediate temperatures $T \sim |E_\delta| + m$, the impurity entropy exhibits a pronounced bump arising from thermal activation of the midgap state. Unlike the simpler YSR I phase discussed previously in Sec.~\ref{sec:YSRI-tba-det}, here the activation energy is effectively shifted by an additional mass gap $m$, associated with the presence of a massive hole excitation. This shift results in negative entropy values near $T \sim |E_\delta|$ before the activation term becomes dominant, reflecting subtle competition between bound-state formation and thermal excitations. 

When $\delta$ is closer to the phase boundary of YSR-I and YSR II phase $(\delta_c=1)$, the impurity is screened in the ground state and unscreened at infinite temperature but exhibits an overshoot above $\ln 2$ at intermediate temperatures, reflecting enhanced impurity fluctuations due to thermal activation of midgap states. In this case, the activation term is enough to overwhelm the contribution from the $\mathcal{T}_1$ tower, resulting in a net positive impurity contribution at all temperature scales.  The insets illustrate the contributions from individual towers, as well as the ensemble-averaged mixing term and the activation term arising from thermal population of the midgap state near $T \sim |E_\delta|+\delta$. These features reveal the interplay between impurity screening and thermal excitation, modulated by $\delta$, which governs the characteristic dips and overshoots in the entropy profile.

In the regime $1 < \delta < \frac{3}{2}$, as previously discussed, the state $\ket{B}$ lies lower in energy because both $\ket{U}$ and $\ket{\tilde{U}}$ involve the presence of a hole excitation. In the state $\ket{B}$, the impurity is screened, whereas in $\ket{U}$ and $\ket{\tilde{U}}$ it remains unscreened. Specifically, in $\ket{U}$, the unscreened impurity and the hole form a triplet multiplet, while in $\ket{\tilde{U}}$, they combine into a singlet. Since the impurity and the hole behave nearly as free particles, their energy difference vanishes in the thermodynamic limit. This is manifested by the fact that the energy of the boundary string $\lambda_{\delta}$ and the higher-order boundary string $\lambda_{\delta,1} = \lambda_{\delta} + i$ are exactly opposite, so that the state $\ket{\tilde{U}}$, constructed from $\ket{U}$ by adding both $\lambda_{\delta}$ and $\lambda_{\delta}+i$ strings, has the same energy as $\ket{U}$.

Thus, the impurity part of the free energy contribution from these three towers constructed on top of states $\ket{B}$, $\ket{U}$, and $\ket{\tilde U}$ reads

\begin{align}
F_{\rm imp}^{(\mathcal{T}_1)} &=m -\frac{T}{4} \int_{-\infty}^\infty \mathrm{d}\lambda \sum_{\upsilon=\pm} \frac{\ln\bigl(1+\eta_3(\lambda)\bigr)}{\cosh\left(\pi(\lambda + i \upsilon (\delta - 1))\right)} + \frac{T}{4} \int_{-\infty}^\infty \mathrm{d}\lambda \sum_{\upsilon=\pm} \frac{\ln\bigl(1+\eta_2(\lambda)\bigr)}{\cosh\left(\pi(\lambda + i \upsilon (\tfrac{3}{2} - \delta))\right)}, \\
F_{\rm imp}^{(\mathcal{T}_2)} &= E_\delta + \frac{T}{4} \int_{-\infty}^\infty \mathrm{d}\lambda \sum_{\upsilon=\pm} \frac{\ln\bigl(1+\eta_1(\lambda)\bigr)}{\cosh\left(\pi(\lambda + i \upsilon (\delta - 1))\right)}, \\
F_{\rm imp}^{(\mathcal{T}_3)} &= m+\frac{T}{4} \int_{-\infty}^\infty \mathrm{d}\lambda \sum_{\upsilon=\pm} \frac{\ln\bigl(1+\eta_2(\lambda)\bigr)}{\cosh\left(\pi(\lambda + i \upsilon (\tfrac{3}{2} - \delta))\right)} + \frac{T}{4} \int_{-\infty}^\infty \mathrm{d}\lambda \sum_{\upsilon=\pm} \frac{\ln\bigl(1+\eta_1(\lambda)\bigr)}{\cosh\left(\pi(\lambda + i \upsilon (\delta - 1))\right)},
\end{align}
where the changes are compared to the discussion in Sec.\ref{sec:YSRIItherm} is the addition of energy $m$ is both $\mathcal{T}_1$ and $\mathcal{T}_2$ due to the addition of a massive hole. However, this change leads to drastic change in the impurity entropy as at low temperature only $\mathcal{T}_2$ contributes to the energy and hence since $S_{\rm imp}^{\mathcal{T}_2}(T=0)=0$, the impurity is screened at low temperature and at infinite temperature as before, all three towers contribute leading to the impurity retaining its free moment entropy value of $\ln 2$. Moreover, at the intermediate temperatures, the thermal population of the midgap states leads to the hump in the entropy at $T\sim  E_\delta$, which leads to an increase in entropy above $\ln 2$ for intermediate temperatures.

As shown in Fig.\ref{fig:del1p3} for the representative value of $\delta=1.3$ in the YSR II phase, the impurity entropy vanishes when $T\to 0$ as impurity is screened in the state $\ket{B}$ which is the ground state and as $T\to\infty$, the impurity is unscreened and hence the impurity entropy reaches $\ln 2$ asymptotically. Moreover, in the intermediate temperatures where the mid-gap states $\ket U$ and $\ket{\tilde U}$ are accessible, the impurity entropy jumps higher than its free moment value of $\ln 2$. 

Just as before, the contributions from each tower at zero temperature are  
\begin{equation}
S_{\rm imp}^{(\mathcal{T}_1)}(T=0) = \ln \frac{3}{2}, \quad
S_{\rm imp}^{(\mathcal{T}_2)}(T=0) = 0, \quad
S_{\rm imp}^{(\mathcal{T}_3)}(T=0) = -\ln 2,
\end{equation}
and in the high-temperature limit, the entropy contributions are given by
\begin{equation}
S_{\rm imp}^{(\mathcal{T}_1)}(T \to \infty) = \ln \frac{4}{3}, \quad
S_{\rm imp}^{(\mathcal{T}_2)}(T \to \infty) = -\ln 2, \quad
S_{\rm imp}^{(\mathcal{T}_3)}(T \to \infty) = -\ln 6.
\end{equation}
  
At zero temperature, only $\mathcal{T}_2$ contributes as $\mathcal{T}_1$ and $\mathcal{T}_3$ has an energy difference of $m-E_\delta>0$. Thus, the zero temperature entropy becomes $S_{\rm imp}(T=0)=\ln(e^{S_{\rm imp}^{(\mathcal{T}_2)}(T=0)})=0$. At infinite temperature, all the towers equally contribute, leading to the infinite temperature total entropy of the impurity to be $S_{\rm imp}(T\to\infty)=\ln\left( e^{S_{\rm imp}^{(\mathcal{T}_1)}(T \to \infty)}+e^{S_{\rm imp}^{(\mathcal{T}_2)}(T \to \infty)}+e^{S_{\rm imp}^{(\mathcal{T}_3)}(T \to \infty)}\right)$. The total entropy due to the three towers for any temperature $T$ can be written as
\begin{equation}
    S_{\rm imp}(T) 
= \ln \left(
    e^{S_{\rm imp}^{(\mathcal{T}_1)}} 
    + e^{S_{\rm imp}^{(\mathcal{T}_2)} - \beta (E_\delta-m)} 
    + e^{S_{\rm imp}^{(\mathcal{T}_3)}}
\right) 
+ \frac{(E_\delta-m)}{T}  
\frac{
    e^{S_{\rm imp}^{(\mathcal{T}_2)} - \beta (E_\delta-m)}
}{
    e^{S_{\rm imp}^{(\mathcal{T}_1)}} 
    + e^{S_{\rm imp}^{(\mathcal{T}_2)} - \beta (E_\delta-m)} 
    + e^{S_{\rm imp}^{(\mathcal{T}_3)}}
},
\end{equation}
where the first term is the weighted ensemble average of the three towers that we call the mixing term, which diverges at $T\to 0$ and asymptotically approaches $\ln 2$ at infinite temperature shown in the left inset, whereas the second term is the activation term which is only non-vanishing when $T\sim E_\delta$ is the contribution from the thermal activation of the midgap state which diverges negatively with the same magnitude as in limit $T\to 0$ and approaches zero in limit $T\to\infty$. Together, these two terms give the total contribution to the impurity entropy, which vanishes in the zero-temperature limit and $\ln 2$ in the infinite-temperature limit, with a pronounced bump overshooting the free moment value of $\ln 2$ in the intermediate temperatures $T\sim E_\delta$.

\begin{figure}[H]
    \centering
    \includegraphics[width=0.75\linewidth]{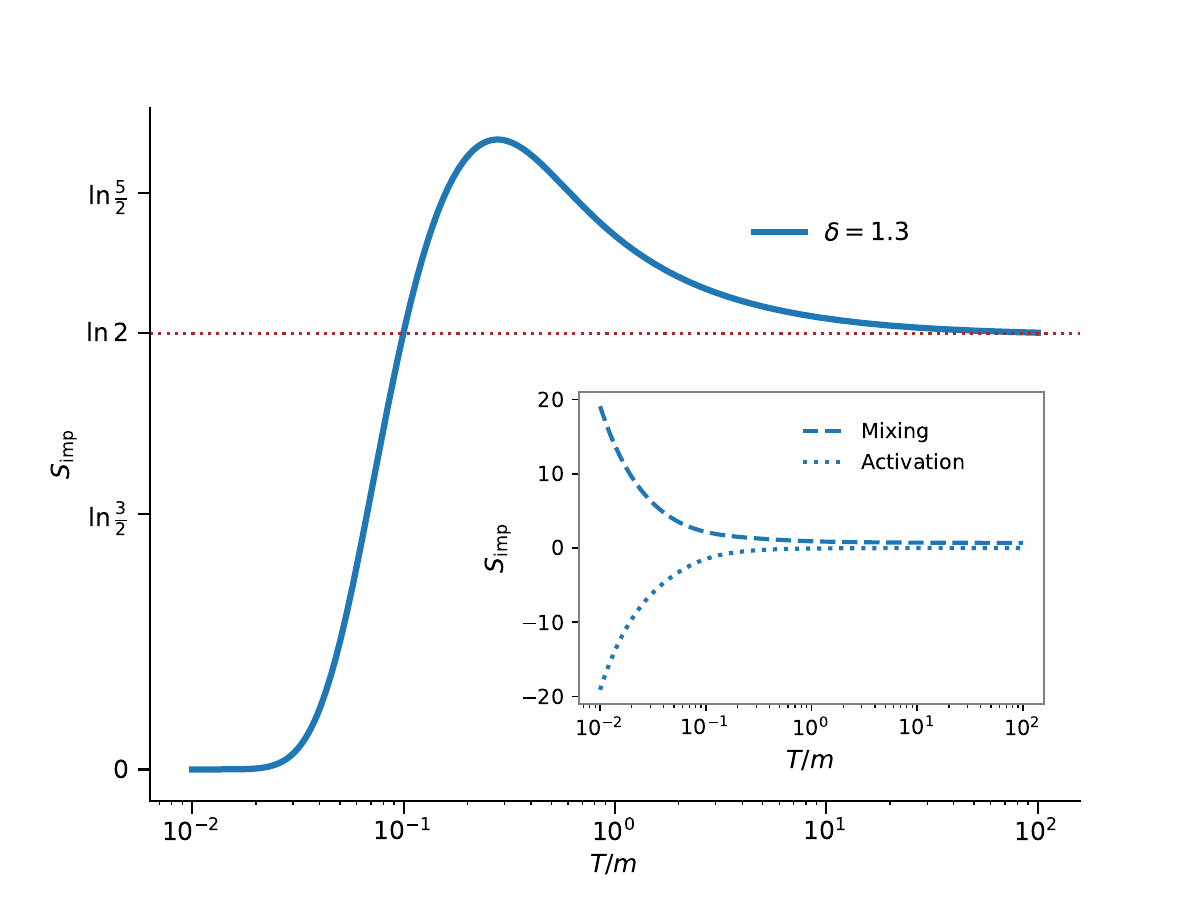}
    \caption{Impurity entropy for representative value of $\delta=1.3$ showing the screening of the impurity in low temperature and asymptotic freedom at high temperature with bump in intermediate temperatures $T\sim E_\delta$. The inset shows the contribution from the ensemble average of the three towers, the mixing term, and the activation term, which is non-vanishing only in narrow windows of the temperature where the midgap state is thermally populated.}
    \label{fig:del1p3pm}
\end{figure}

\section{Lattice model}
	
	Consider the one-dimensional attractive Hubbard chain with a spin-$\frac{1}{2}$ impurity coupled to the site $j=1$ described the lattice Hamiltonian
	\begin{equation}
		H = -t \sum_{j=1}^{L-1}\sum_{\sigma=\uparrow,\downarrow} \bigl( c_{j\sigma}^\dagger c_{j+1\sigma} + \text{h.c.} \bigr)
		- U \sum_{j=1}^L n_{j\uparrow} n_{j\downarrow} + J_{\mathrm{imp}} \bm{S}_{\mathrm{imp}} \cdot \bm{s}_1,
		\label{hubb+imp}
	\end{equation}
	where $U>0$ (attractive), $\bm{s}_j = \frac{1}{2} c^\dagger_{j\alpha} \bm{\sigma}_{\alpha\beta} c_{j\beta}$ and $\bm S_{\rm imp}$ is the spin operator of the localized impurity. When $J_{\rm imp}=0$, the model is integrable and can be solved exactly via the Bethe Ansatz~\cite{lieb1968absence}. However, in the presence of the impurity, the model may not be integrable. However, the behavior of the impurity can be studied analytically by looking at the low-energy theory constructed by linearizing the spectrum around the Fermi sea. To do so, we employ non-Abelian bosonization, where the spin sector is described by an SU(2)$_1$ Wess-Zumino-Witten (WZW) model. The fermion fields map to
	\begin{equation}
		\psi_{R\sigma} \sim \frac{1}{\sqrt{2\pi}} e^{i\sqrt{4\pi}\phi_R^c} g_{R\sigma}, \quad
		\psi_{L\sigma} \sim \frac{1}{\sqrt{2\pi}} e^{-i\sqrt{4\pi}\phi_L^c} g_{L\sigma},
	\end{equation}
	where $\phi^c$ is the charge boson, and $g$ is an SU(2)-valued matrix field. The currents are:
	\begin{equation}
		\mathcal  J_L^a = -\frac{i}{2\pi} \mathrm{tr}(\sigma^a g_L^{-1}\partial g_L), \quad
		\mathcal  J_R^a = -\frac{i}{2\pi} \mathrm{tr}(\sigma^a g_R^{-1}\bar{\partial} g_R)
	\end{equation}
	The free Hamiltonian is:
	\begin{equation}
		H_0 = \frac{2\pi}{3} \int dx \left( \mathcal J_L  ^a\mathcal{J}_L^a + \mathcal J_R^a \mathcal{J}_R^a \right) + \frac{1}{2} \int dx \left[ (\partial_x \phi^c)^2 + (\partial_x \theta^c)^2 \right],
	\end{equation}
	where $\theta^c$ is the dual field to $\phi^c$. The bulk interaction becomes
	\begin{equation}
		H_{\rm int} = \frac{g}{\pi} \int \mathrm dx~ \mathcal{J}_R^a \mathcal{J}_L^a.
	\end{equation}
	The boundary Kondo interaction
	\begin{equation}
		H_K = J ~ {S}_{\rm imp}^a \left( \mathcal{J}_R^a(0) + \mathcal{J}_L^a(0) \right)
	\end{equation}
	couple the impurity to the spin current at the boundary.  Here $g\propto\frac{|U|}{t}$ and $J\propto J_{\rm imp}$.  Notice that $H_0+H_{\rm int}+H_K$ is precisely the Gross-Neveu model with magnetic impurity at the boundary that we considered in the main text, which is integrable. 
	
	Notice that in the absence of the Kondo term $J=0$, the bulk perturbation $H_{\rm int}$ is a perturbation with scaling dimension $\Delta=2$, which is classically marginal, and hence the beta function vanishes at tree level. However, the conformal fixed point is destabilized by a marginally relevant perturbation, the product of left- and right-moving spin currents, $\mathcal{J}_L^a\mathcal{J}_R^a$, as at one-loop order, the operator product expansion (OPE) of the interaction with itself generates a logarithmic ultraviolet divergence. This divergence manifests as a nontrivial renormalization of the coupling constant, yielding a one-loop beta function of the form $\beta(g)=-\frac{2}{\pi}g^2$, which causes the coupling constant $g$ to increase as the energy scale is lowered, driving the system toward a strong coupling regime in the infrared limit.  To define the renormalized theory consistently, an ultraviolet cutoff $D$ is introduced, and the bare coupling $g(D)$ flows such that the dynamically generated mass scale 
	\begin{equation}
		m=De^{-\frac{1}{g(D)}}
	\end{equation}
	remains fixed as $D\to\infty$. This phenomenon of dimensional transmutation leads to the opening of a spin gap, dynamically gapping the spin excitations and fundamentally altering the low-energy physics by breaking conformal invariance.

	Likewise, in the absence of the bulk interaction, i.e., when $g=0$, the boundary Kondo perturbation has scaling dimension $\Delta=1$, which matches with the space-time dimension at the boundary, rendering the interaction classically marginal. Just as in the case of bulk interaction, the one-loop beta function of the boundary coupling becomes $\beta(J)=-\frac{2}{\pi}J^2$. However, in the presence of the bulk interaction, the scaling dimension of the bulk changes from $\Delta=1$ to $\Delta=1+g$ to the first order in $g$, and hence at the tree level, the beta function gets an additional  $\frac{2}{\pi} gJ$ contribution such that the beta function of the boundary coupling $J$ in the presence of the bulk coupling $g$ becomes 
	\begin{equation}
		\beta(J)=-\frac{2}{\pi}J(J-g)
		\label{betaJeqn}
	\end{equation}
	Thus, only when $J>g$, the beta function remains negative, and hence boundary coupling grows as the energy scale is lowered. But when $g>J$, the beta function reverses sign and hence the boundary coupling flows to zero, thereby leaving the impurity effectively decoupled from the impurity.

	While perturbation theory captures the Kondo and local moment phases, it overlooks an intermediate phase that occurs when $J\sim g$, featuring an exponentially localized midgap state at the edge that screens the impurity.  However, a simple semiclassical analysis can reveal those states. To do so, we approximate the gapped Hamiltonian in the bulk by a gapped Dirac Bogoliubov–de Gennes (BdG) Hamiltonian of the form
	\begin{equation}
		\mathcal{H}_{\mathrm{bulk}} = 
		\begin{pmatrix}
			-i \partial_x & m & 0 & 0 \\
			m & i \partial_x & 0 & 0 \\
			0 & 0 & -i \partial_x & m \\
			0 & 0 & m & i \partial_x
		\end{pmatrix},
	\end{equation}
	acting on the spinor
	$\Psi(x) = (\psi_{R\uparrow}, \psi_{L\downarrow}^\dagger, \psi_{R\downarrow}, \psi_{L\uparrow}^\dagger)^T$.
	with eigenvalues
	\begin{equation}
		E = \pm \sqrt{k^2 + m^2}.
	\end{equation}
	
	The local Green’s function at the impurity site is given by
	\begin{equation}
		G_{0,\sigma}(\omega) = -\frac{1}{2 \sqrt{m^2 - \omega^2}} \left( \omega \tau_0 + m \tau_x \right),
	\end{equation}
	where $\tau_i$ are Pauli matrices in particle-hole space.
	
	A classical magnetic impurity polarized along the $z$-axis couples as
	\begin{equation}
		\Sigma = - J S_{\mathrm{imp}}^z \sigma_z \otimes \tau_0,
	\end{equation}

	The impurity Green’s function satisfies the Dyson equation
	\begin{equation}
		G_\sigma(\omega) = \left[ G_{0,\sigma}^{-1}(\omega) - \Sigma_\sigma \right]^{-1}.
	\end{equation}
	
	The in-gap bound state energies $|\omega| < m$ satisfy
	\begin{equation}
		\det \left[ G_{0,\sigma}^{-1}(\epsilon_0) - \Sigma_\sigma \right] = 0,
	\end{equation}
	which leads to the YSR energy
	\begin{equation}
		\epsilon_{\mathrm{YSR}} = \pm m \frac{1 - \alpha^2}{1 + \alpha^2}, \quad \alpha = \frac{J S_{\mathrm{imp}}^z}{2}.
	\end{equation}
	
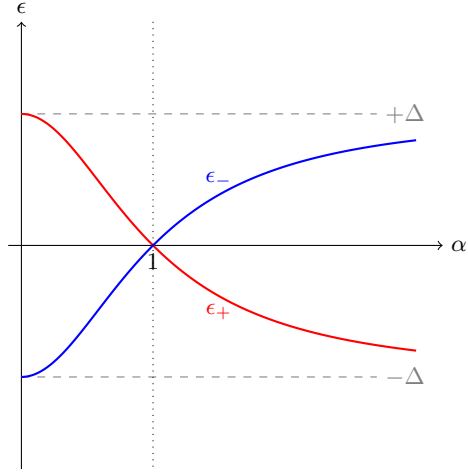
\begin{figure}[H]
		\centering
		\begin{tikzpicture}[scale=1.75]
			\draw[->] (-0.1,0) -- (3.2,0) node[right] {$\alpha$};
			\draw[->] (0,-1.7) -- (0,1.7) node[above] {$\epsilon$};
			
			\draw[dashed, gray] (0,1) -- (2.7,1) node[right] {$+\Delta$};
			\draw[dashed, gray] (0,-1) -- (2.7,-1) node[right] {$-\Delta$};
			
			\draw[thick, red, domain=0:3, smooth, samples=100] plot (\x, {(1 - \x*\x)/(1 + \x*\x)});
			\draw[thick, blue, domain=0:3, smooth, samples=100] plot (\x, {-(1 - \x*\x)/(1 + \x*\x)});
			
			\node[blue] at (1.5,0.5) {$\epsilon_-$};
			\node[red] at (1.5,-0.5) {$\epsilon_+$};
			
			\draw[dotted] (1,1.7) -- (1,-1.7);
			\node[below] at (1,0) {$1$};
		\end{tikzpicture}
		\caption{YSR bound state energies $\epsilon_{\pm}/m = \pm \frac{1 - \alpha^2}{1 + \alpha^2}$ plotted for $\alpha \geq 0$. The bound states move from the gap edges at $\alpha = 0$ to zero energy at $\alpha = 1$.}
		\label{fig:YSR-cs}
	\end{figure}

	As shown in Fig.~\ref{fig:YSR-cs}, this analysis shows that the YSR mid-gap states exist throughout the phase space and for $\alpha=1$, there is a level crossing such that the model undergoes a quantum phase transition at $\alpha=1$. However, the full quantum solution shows that the quantum fluctuation restricts the YSR phase to a small regime when $g\sim J$, but for $g\ll J$, the impurity is unscreened, and for $J\ll g$, the impurity is screened by a multiparticle Kondo cloud. 
	
	The goal of this section is to focus on the YSR regime directly within the lattice Hamiltonian defined in Eq.~\eqref{hubb+imp} for finite-size systems. We study this system using exact diagonalization. First, we shall identify the distinct YSR(I) and YSR(II) regimes, compute the corresponding thermodynamic properties, and demonstrate that the impurity entropy exhibits an overshoot beyond its free moment value of $\ln 2$ at intermediate temperatures, consistent with the exact solution presented in the main text.
	
	\subsection{Numerical solution}
	We begin by studying the finite-size lattice Hamiltonian defined by Eq.~\eqref{hubb+imp}, describing a spin-$\frac{1}{2}$ impurity coupled to an interacting electron bath of $N-1 = 7$ sites. The full Hamiltonian is given explicitly by
	\begin{equation}
		\hat{H} = -t \sum_{j=1}^{6} \sum_{\sigma = \uparrow, \downarrow} \left( c_{j,\sigma}^\dagger c_{j+1,\sigma} + \text{h.c.} \right)
		+ U \sum_{j=1}^7 \hat{n}_{j,\uparrow} \hat{n}_{j,\downarrow}
		+ J_{\mathrm{imp}} \left( S_0^+ s_1^- + S_0^- s_1^+ + 2 S_0^z s_1^z \right),
		\label{ham-FS}
	\end{equation}
	where site $0$ is the impurity and sites $1 \ldots 7$ form the bulk. The electron hopping amplitude is set to $t = 1$, and the bulk onsite interaction is fixed at $U = -5$, corresponding to an attractive Hubbard interaction.
	
\begin{figure}[H]
		\centering
		\includegraphics[width=0.48\linewidth]{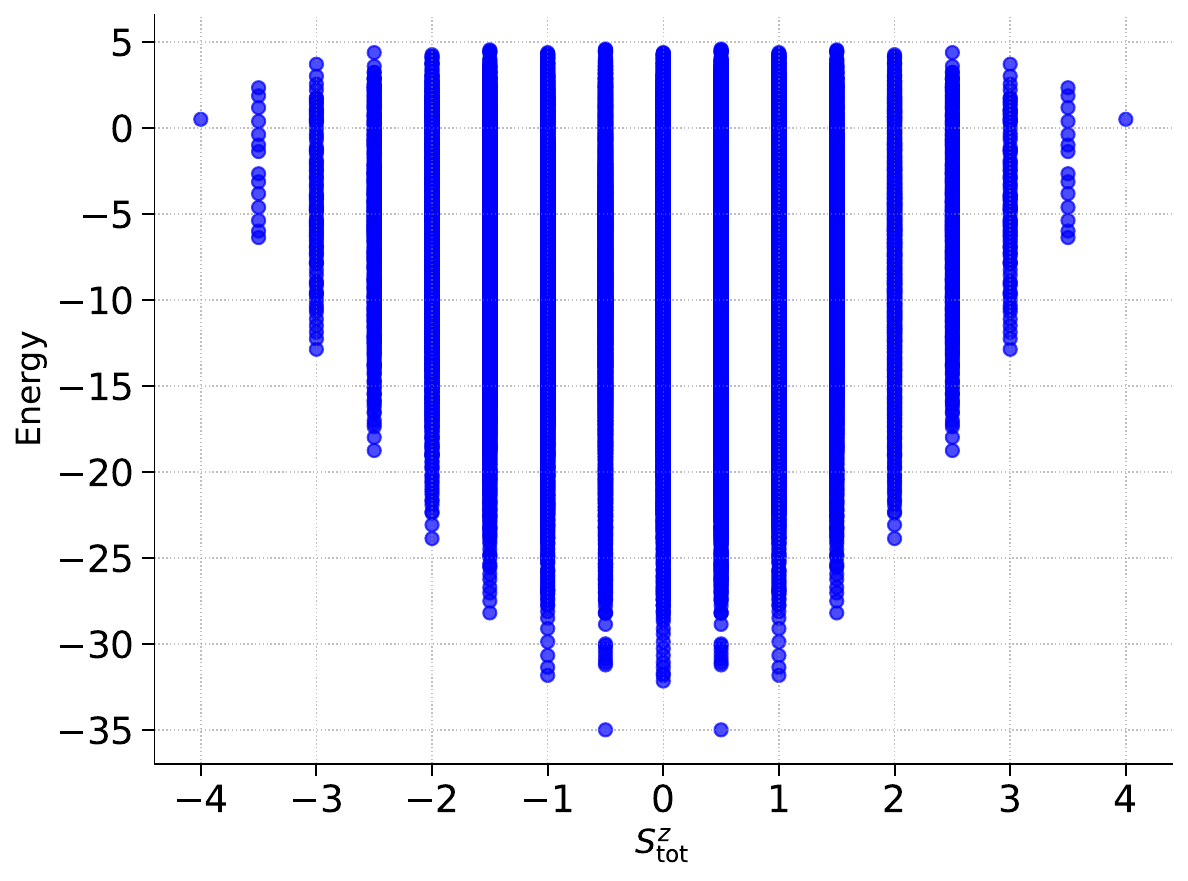}
		\includegraphics[width=0.48\linewidth]{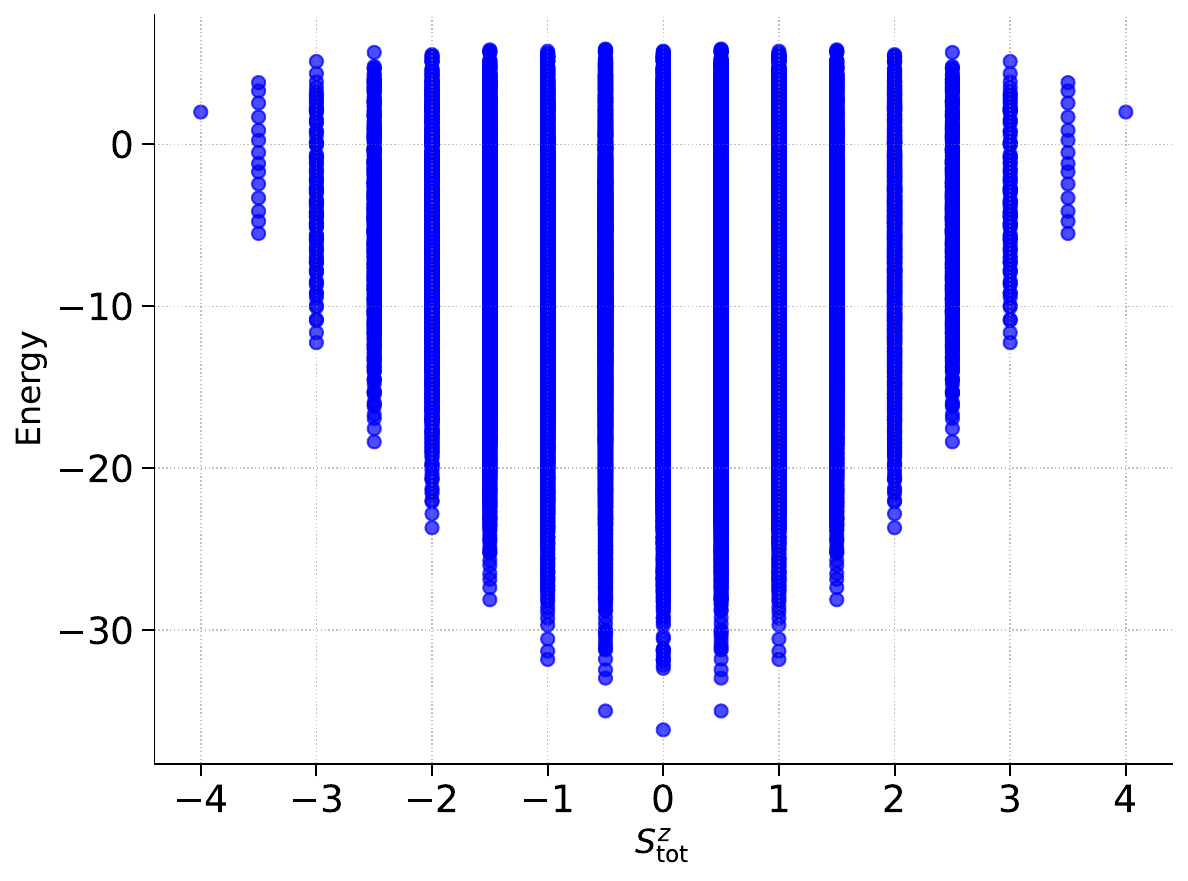}
		\caption{
			The entire spectrum of the Hamiltonian Eq.~\eqref{ham-FS} is computed via exact diagonalization with onsite interaction $U = -5$ and hopping $t = 1$.  
			Left panel: $J_{\mathrm{imp}} = 0.5$, impurity remains unscreened with a two-fold degenerate ground state at $S^z = \pm \frac{1}{2}$.  
			Right panel: $J_{\mathrm{imp}} = 2.0$, impurity is screened, resulting in a unique singlet ground state, and there are two-fold degenerate mid-gap states with unscreened impurity.
		}
		\label{fig:spectrum-ED}
	\end{figure}

	We first analyze the full system for two impurity-bath exchange coupling values:
	\begin{itemize}
		\item $J_{\mathrm{imp}} = 0.5$, where the impurity remains unscreened and hence the ground state is two-fold degenerate,
		\item $J_{\mathrm{imp}} = 2.0$, where the impurity becomes screened by the conduction electrons, resulting in a  unique singlet ground state,
	\end{itemize}
	as shown in Fig.~\ref{fig:spectrum-ED}.
	
\begin{figure}[H]
		\centering
		\includegraphics[width=0.5\linewidth]{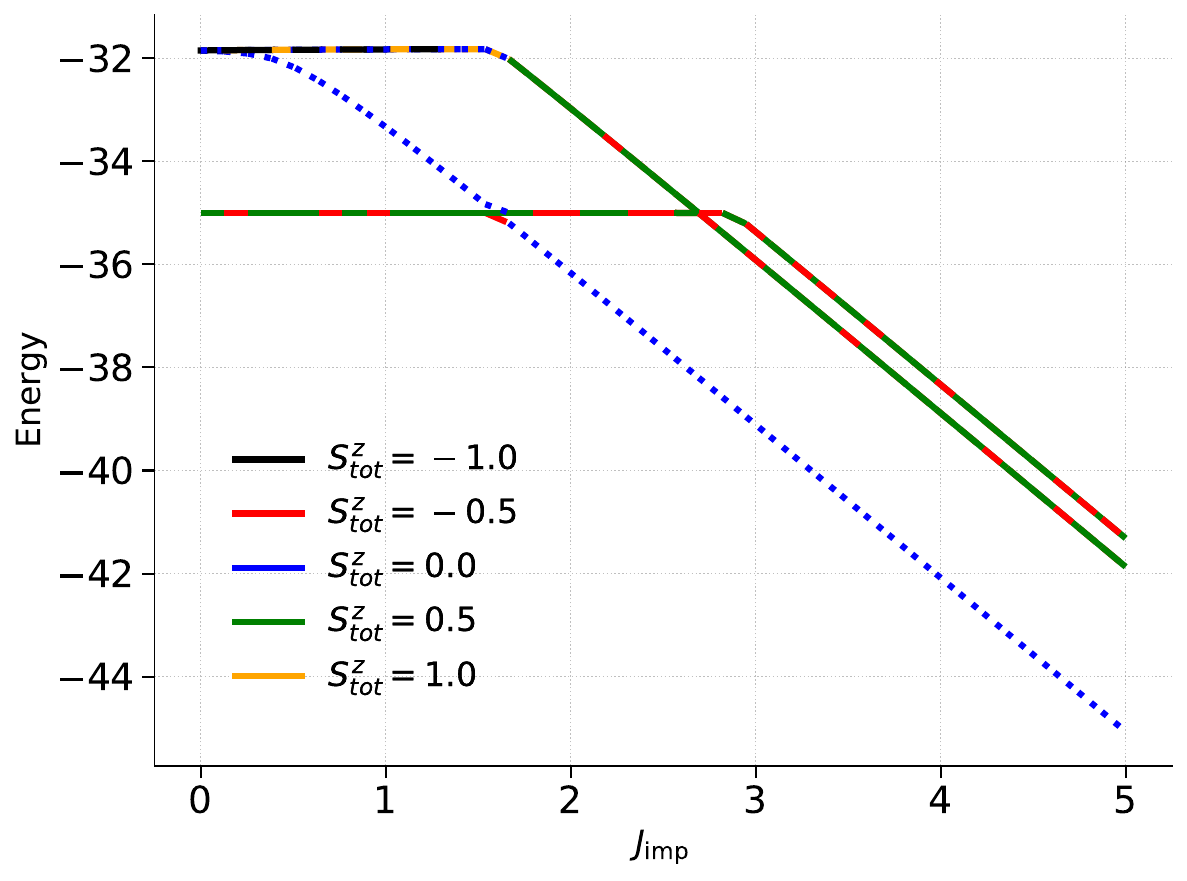}
		\caption{Low-lying energy spectrum of the finite-size lattice Hamiltonian Eq.\eqref{ham-FS} as a function of the impurity coupling $J_{\mathrm{imp}}$ with fixed parameters $U = -5$ and $t = 1$.  
			The five lowest excited states are shown, with colors indicating the total spin projection $S^z_{\mathrm{tot}}$ of each state.  
			A clear level-crossing transition occurs near $J_{\mathrm{imp}} \approx 1.6$, reflecting the change in ground state from a doublet ($S^z=\pm \frac{1}{2}$) to a singlet $S^z=0$.}
		\label{fig:level-crossing}
	\end{figure}

	To understand the formation of the YSR state, we examine the evolution of the low-lying energy spectrum by fixing $U = -5$ and $t = 1$ while continuously varying the impurity coupling $J_{\mathrm{imp}}$. As illustrated in Fig.~\ref{fig:level-crossing}, the system undergoes a level-crossing transition near $J_{\mathrm{imp}}^{c} \approx 1.6$, marking a qualitative change in the ground state structure. For couplings $J_{\mathrm{imp}} < J_{\mathrm{imp}}^{c}$, the impurity remains unscreened, whereas for $J_{\mathrm{imp}} > J_{\mathrm{imp}}^{c}$, it becomes screened by the conduction electrons as shown by the colors indicating the total soin of the state in Fig.~\ref{fig:level-crossing}.
	
\begin{figure}[H]
		\centering
		\includegraphics[width=0.48\linewidth]{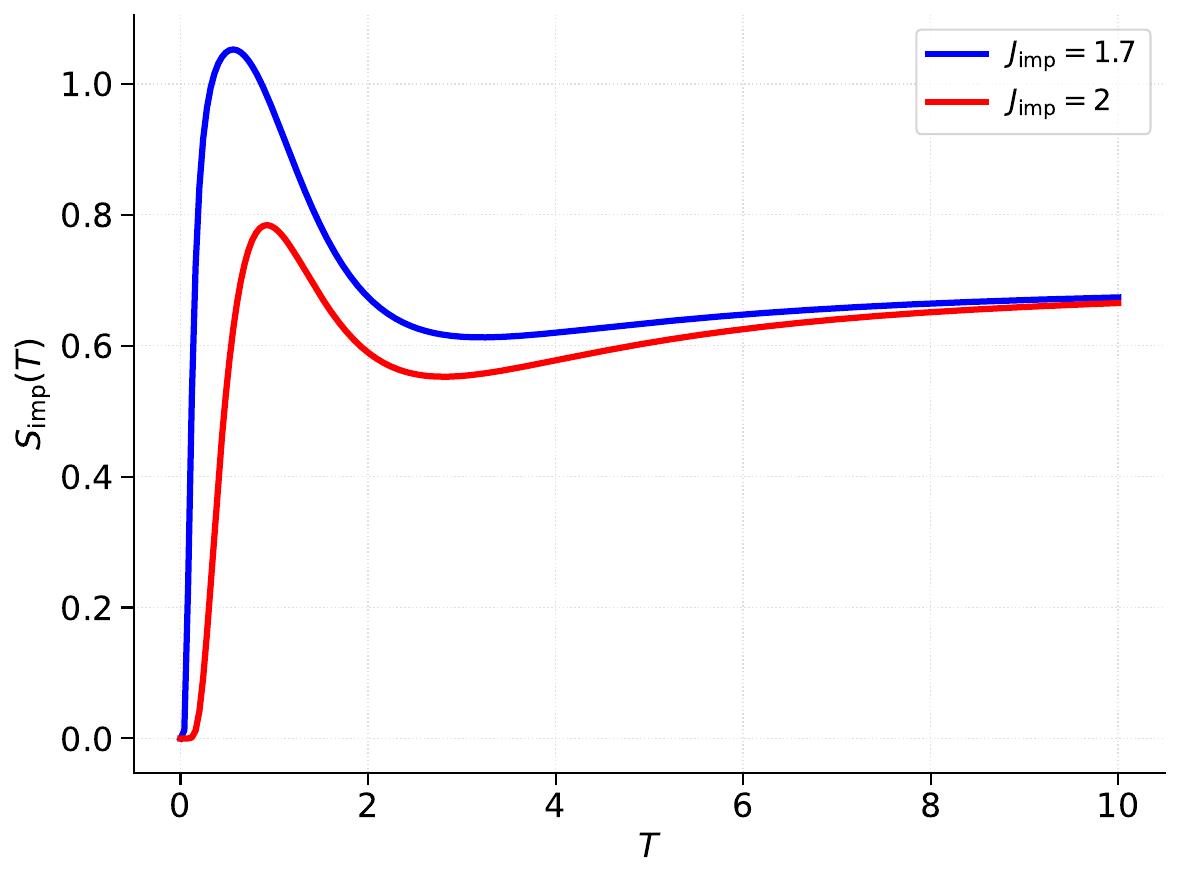}
		\includegraphics[width=0.48\linewidth]{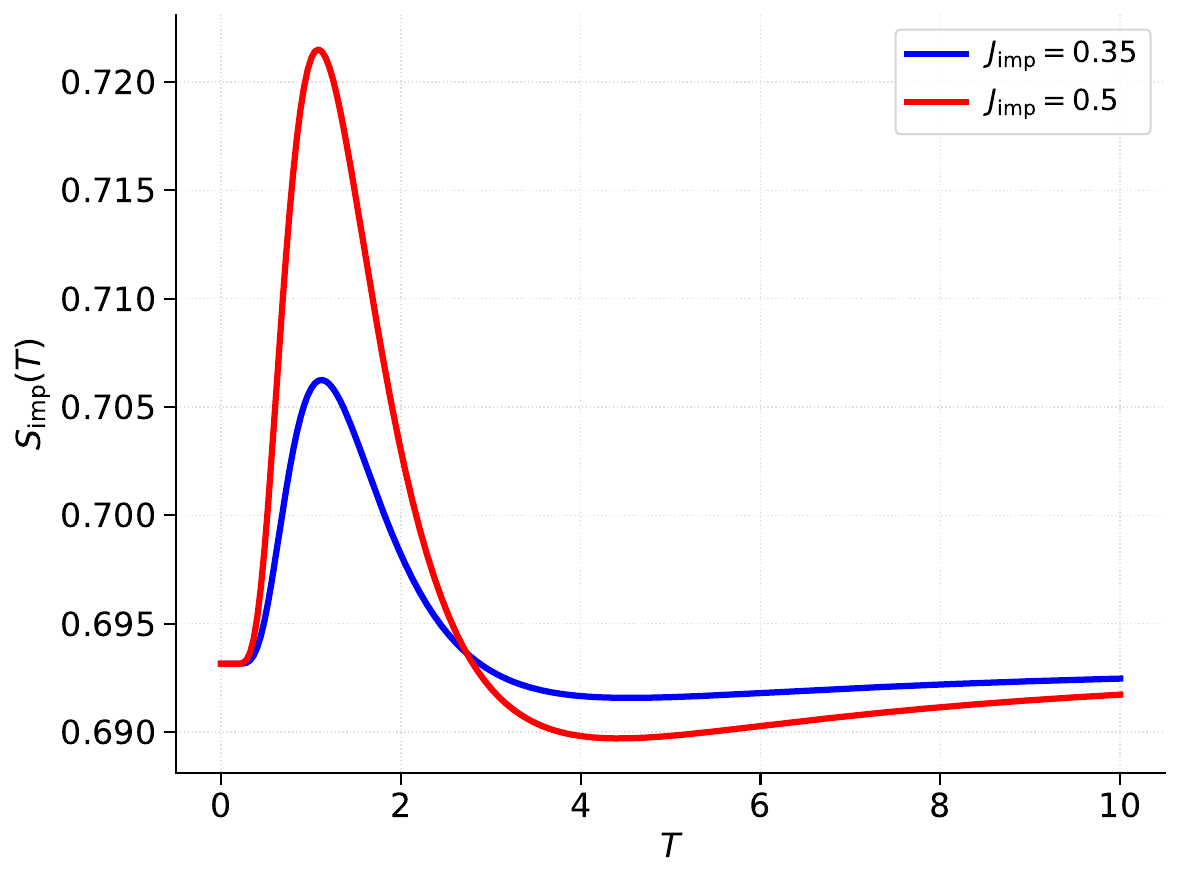}
		\caption{
			\textbf{Left panel:} In the YSR I phase, the impurity is screened at low temperature and unscreened at high temperature, resulting in an impurity entropy that vanishes as $T \to 0$ and approaches $\ln 2$ as $T \to \infty$. Unlike the conventional Kondo effect, the entropy exhibits a pronounced nonmonotonic behavior, even peaking above the free-moment value of $\ln 2$ at intermediate temperatures where the midgap state is thermally accessible, as shown for the representative values of $J_{\rm imp}=\{1,7,2\}$.
			\textbf{Right panel:} In the YSR II phase, the impurity remains unscreened at both low and high temperatures, thereby retaining the free-moment entropy value of $\ln 2$ in these limits. At intermediate temperatures, where the midgap state is thermally populated, the entropy overshoots $\ln 2$, as shown for the representative values of $J_{\rm imp}=\{0.35,0.5\}$.}
		\label{fig:impurity_entropy}
	\end{figure}

	Having studied the low-lying energy spectrum, we can now identify the YSR(II) and YSR(I) regimes. We proceed by computing the impurity entropy for representative values of the impurity coupling: $J_{\mathrm{imp}} = 0.5$ within the YSR(II) regime, where the impurity remains unscreened in the ground state, and $J_{\mathrm{imp}} = 2$ in the YSR(I) regime, where the impurity is screened. As shown in Fig.~\ref{fig:impurity_entropy}, these cases exhibit distinct thermodynamic behavior reflecting their different ground state properties. As discussed in the main text, in the YSR (I) regime, the impurity is screened at low temperature and becomes asymptotically free at high temperature. Consequently, the impurity entropy flows from zero at zero temperature to the free-moment value $\ln 2$ at infinite temperature, exhibiting a pronounced hump at intermediate temperatures. In contrast, in the YSR (II) regime, the impurity remains unscreened at both zero and infinite temperatures, thus retaining the free moment entropy $\ln 2$ at these extremes. However, at intermediate temperatures where the mid-gap state becomes thermally accessible, the impurity entropy overshoots the free-moment value $\ln 2$. We illustrate these behaviors for representative values $J_{\mathrm{imp}} = 1.7$ and $J_{\mathrm{imp}} = 2.0$ in the YSR (I) phase, and $J_{\mathrm{imp}} = 0.35$ and $J_{\mathrm{imp}} = 0.5$ in the YSR (II) phase in Fig.~\ref{fig:impurity_entropy}.

\end{document}